\documentclass[aps,prb,showpacs,floatfix,twocolumn,superscriptaddress]{revtex4-1}
\usepackage{graphicx}
\usepackage{hyper ref}
\usepackage{bm,amsmath}
\usepackage{dcolumn}
\newcommand{\be}{\begin{equation}}
\newcommand{\ee}{\end{equation}}
\newcommand{\bea}{\begin{eqnarray}}
\newcommand{\eea}{\end{eqnarray}}
\newcommand{\la}{\langle}
\newcommand{\ra}{\rangle}

\newcommand{\bfS}{{\bf S}}

\newcommand{\half}{\frac{1}{2}}
\graphicspath{{fig_tpc/}}

\begin{document}

\title{Typicality at quantum-critical points}

\author{Lu Liu}
\affiliation{Department of Physics, Beijing Normal University, Beijing 100875, China }

\author{Anders W. Sandvik}
\email{sandvik@bu.edu}
\affiliation{Department of Physics, Boston University, 590 Commonwealth Avenue, Boston, Massachusetts 02215}
\affiliation{Beijing National Laboratory for Condensed Matter Physics and Institute of Physics, Chinese Academy of Sciences, Beijing 100190, China}

\author{Wenan Guo}
\email{waguo@bnu.edu.cn}
\affiliation{Department of Physics, Beijing Normal University, Beijing 100875, China }

\begin{abstract}
We discuss the concept of typicality of quantum states at quantum-critical points, using projector Monte Carlo simulations
of an $S=\half$ bilayer Heisenberg antiferromagnet as an illustration. With the projection (imaginary) time $\tau$ scaled as $\tau =aL^z$,
$L$ being the system length and $z$ the dynamic critical exponent (which takes the value $z=1$ in the bilayer model studied here),
a critical point can be identified which asymptotically flows
to the correct location and universality class with increasing $L$, independently of the prefactor $a$ and the initial state.
Varying the proportionality factor $a$ and the initial state only changes the cross-over behavior into the asymptotic large-$L$
behavior. In some cases, choosing an optimal factor $a$ may also lead to the vanishing of the leading finite-size corrections. The observation
of typicality can be used to speed up simulations of quantum criticality, not only within the Monte Carlo approach but also with other
numerical methods where imaginary-time evolution is employed, e.g., tensor network states, as it is not necessary to evolve
fully to the ground state but only for sufficiently long times to reach the typicality regime.
\end{abstract}
\maketitle

\section{Introduction}

Typicality in quantum many-body physics refers to the emergence in large systems of typical properties
of arbitrary pure states that depend only on global control variables such as the energy.\cite{tpc1,tpc2,tpc3}  If an observable is typical,
details of the initial state preparation in an experiment do not matter. The perhaps most striking example of typicality
is the eigenstate thermalization hypothesis (ETH),\cite{eth,tpq1,tpq2} according to which a single eigenstate suffices to characterize
the properties of a statistical ensemble of states at the temperature corresponding to the energy.  The ETH and other
manifestations of typicality are now believed to hold generically, but with important exceptions in, e.g., integrable systems. \cite{Deutsch, noeth}

The concept of typicality is at the heart of fundamentally understanding how macroscopic properties
emerge from the microscopic scale in quantum systems. It is also of practical interest experimentally,
especially as deviations from typicality become important in nano-scale systems.
With the increasing importance of numerical simulations in quantum many-body physics for systems that are analytically
intractable, the issue of typicality is also of key importance both in interpreting simulation results and for setting up
simulation protocols.

In numerical studies of finite-temperature properties using eigenstate-based method (exact diagonalization or Lanczos calculations),
it has for some time been known that the trace over states needed in a quantum mechanical expectation value of some
observable $A$ at inverse temperature $\beta$,
\be
\la A(\beta) \ra= Z_\beta^{-1} {\rm Tr} \{ A {\rm e}^{-\beta H}\},~~~~Z_\beta={\rm Tr}\{{\rm e}^{-\beta H}\},
\ee
does not have to be evaluated completely; it is normally sufficient to average over a small number of states (or even a single state).
\cite{jaklic94,jaklic00}
This observation was made more precise and was utilized as a way to optimize the averaging procedure using ``minimally entangled states''
in $T>0$ calculations with the density-matrix renormalization group method.\cite{White,Garnerone}
Here we will show that the concept of typicality can also be used in studies of quantum phase transitions, where the focus is on
grounds states. Though the target of a calculation in this case is a specific eigenstate, we will show that the typical critical
scaling properties also emerge in classes of states that resemble thermal mixed states.

\subsection{Typicality in imaginary time evolution}

We work in the context of projector quantum Monte Carlo (PQMC) simulations,\cite{pqmcliang,pqmc} where, given an essentially arbitrary
initial state $|\Psi_0\ra$ (often called a ``trial state'', though the term is somewhat misleading), the ground state can be found by
time evolution in imaginary time,
\be
|\Psi(\tau)\ra =U(\tau) |\Psi(0)\ra,
\label{UPsi}
\ee
with $U=\exp(-\tau H)$ and $\tau$ sufficiently large in a way that we will make more precise. The expectation of an observable $A$
in the projected state is calculated as
\be
\la A(\tau) \ra=Z_\tau^{-1}{\la \Psi(\tau) |A|\Psi(\tau)\ra},~~~~Z_\tau=\la \Psi(\tau) |\Psi(\tau) \ra.
\label{expA}
\ee
For a finite system of linear size $L$, the ground state expectation value $\la 0|A|0\ra = \la A(\tau \to \infty) \ra$ can always be
obtained (provided that $|\Psi\ra$ has some overlap with it) to arbitrary precision by using some large value of $\tau$.
To be systematic, one can, for example, double $\tau$ in a series of calculations until the results converge within statistical errors.
This is similar to $T>0$ methods, such as the stochastic series expansion (SSE) quantum Monte Carlo (QMC) method, \cite{sse,qspin} applied
for a series of inverse temperatures $\beta=1/T$, e.g., $\beta_n=2^n$, to make sure that the groundstate properties emerge
as $n$ is increased (see, e.g., Ref.~\onlinecite{wang10} for a case where extremely low temperatures were reached this way).

Although in some cases one would like to reach the true ground state, in studies of continuous quantum critical points it is well known
that this is not necessary. Instead, if the dynamic exponent $z$ is known, one can choose $\beta=aL^z$, where $z$ is the dynamic
critical exponent, with an arbitrary proportionality factor $a$. This scaling of $\beta$ removes the dependence on $\beta$ from the
scaling function, and one can then study finite-size scaling of computed quantities only in the spatial size $L$. This is often a
more practical (less demanding of computer resources) alternative to eliminating the $\beta$ dependence by effectively taking the limit
$\beta \to \infty$ when studying the true ground state.

In analogy with $T>0$ simulations of criticality, we will here carry out PQMC calculations with the projection time scaled as $\tau= aL^z$
for a quantum-critical system. This has certainly been done before, under the intuitively clear notion that the temporal boundary conditions
should not matter and one can proceed exactly as in $T>0$ calculations; see, e.g., the recent work in Ref.~\onlinecite{shu17}.
However, the freedom to choose the trial state is an aspect of the problem not present in $T>0$ simulations, where one always has
periodic imaginary time boundaries. To our knowledge, systematic studies of the role of the trial state and how the typicality
emerges irrespective of it and the scale factor in the projection time has not been studied systematically before. We will here demonstrate
this independence of the asymptotic scaling behaviors on $a$ as well as the trial state $|\Psi(0)\ra$, by choosing a range of different
trial states and factors. Our results in all cases demonstrate quantum-critical typicality of imaginary-time evolved states. On a practical
level, this can save time in PQMC simulations, as the simulation time typically scales linearly with $\tau$ and reaching the true ground
state to within small error bars may require very large $\tau$ and detailed tests to check for convergence. Instead scaling $\tau$ as $aL$
with a small factor $a$ can then save significant computer time. However, as one may expect, if $a$ is too small, very large system sizes
are required for the asymptotic critical behavior to set in. We will observe how the cross-over depends on $a$ and the trial state and
also discuss possible elimination of the leading scaling corrections by identifying an optimal value of $a$. We will also make other
interesting observations on the role of the trial state.

\subsection{Critical Bilayer Heisenberg Model}

\begin{figure}
	\includegraphics[width=7.5cm]{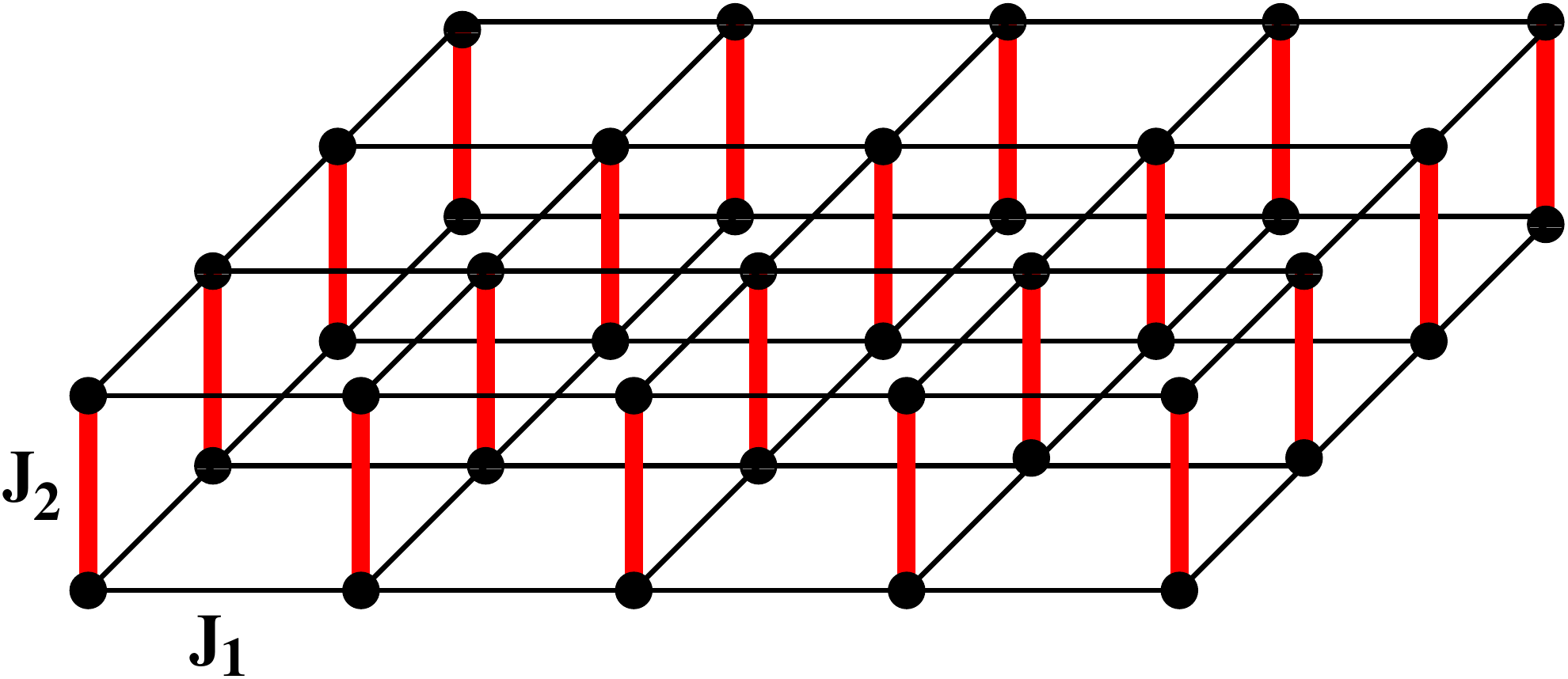}
	\caption{The bilayer Heisenberg model with two different exchange constants between nearest-neighbor $S=1/2$ spins;
          $J_1$ within the individual layers and $J_2$ between the layers.}
	\label{dblh}
\end{figure}

To validate the above typicality hypothesis, we use the $S=1/2$ Heisenberg model on a symmetric bilayer,\cite{blayer1,blayer2,blayer3}
illustrated in Fig. \ref{dblh}, as a concrete example. The Hamiltonian of the model is given by
\be
H=J_1 \sum_{\la i,j \ra} (\bfS_{1i} \cdot \bfS_{1j} + \bfS_{2i}\cdot \bfS_{2j})
+J_{2}\sum_{i=1}^{L^2} \bfS_{1i}\cdot \bfS_{2i},
\label{H_blyer}
\ee
where $\bfS_{ai}$ is a spin $S=1/2$ operator at site $i$ of layer $a=1, 2$ and $\la i, j\ra$ denotes a nearest-neighbor pair of spins on
the $L\times L$ square lattice. Periodic boundary conditions are applied. The coupling constants $J_1$ and $J_{2}$ are antiferromagnetic
(positive). Only even sizes $L$ are considered in our study to avoid frustration due to the periodic boundary conditions imposed.

The bilayer model realizes a quantum phase transition from N\'eel order for small $g=J_2/J_1$ (e.g., for $g=0$ the system consists
of two decoupled 2D Heisenberg layers, for which the long-range N\'eel order is well understood and quantified \cite{manousakis91})
to a quantum paramagnet for large values. For $g \to \infty$ the ground state is simply a product of singlets on the bond between
the layers and there is a spin gap $\Delta=J_2$. This gap closes at $g_c$, and long-range N\'eel order forms continuously for $g<g_c$.
By symmetry, this $T=0$ transition in 2+1 dimensions (two space dimensions and one time dimension) should belong to the universality
class of the finite-T transition of the three-dimensional (3D) classical Heisenberg model, or O(3) model,\cite{CHN,CSY,sachdev} provided
that the Lorentz invariance is emergent when $L \to \infty$ (i.e., the dynamic exponent takes the value $z=1$).

The N\'eel--paramagnetic transition has been studied intensely, with the bilayer model and several other cases of dimerization; for
a review, see Ref.~\onlinecite{qspin}.  Among the more precise studies of the bilayer, in Ref.~\onlinecite{dbAFH} SSE calculations were used
to determine the critical coupling ratio as $g_c=2.5220(1)$
(where here and henceforth the number within parenthesis indicates the statistical error of the preceding digit) and the correlation length
exponent was found to be $\nu=0.7106(9)$, which is in agreement with the 3D classical Heisenberg exponent $\nu = 0.7112(5)$  obtained in
a high-precision study of the classical 3D $\phi^4$ model.\cite{Vicari} This exponent is also consistent with that of the 2D columnar dimerized
quantum Heisenberg model.\cite{clmnH} Initial disagreement with O(3) universality for the case of staggered dimers \cite{Janke1} have now been
attributed to strong corrections to scaling.\cite{FJJiang,Fritz12,NSMa} Thus, there is little doubt that all these dimerized models belong to the
same standard O(3) universality class. Here our goal is not to reconfirm this or to obtain more precise exponents, but to convincingly demonstrate
that typicality at the quantum critical fluctuations holds, in the precise sense that the same O(3) exponents are produced asymptotically
(for sufficiently large system size) with different prefactors $a \le 1$ in the scaling of $\tau$ with $L$ in PQMC simulations and with
a set of completely different trial states in Eq.~(\ref{expA}). We only assume that the dynamical exponent $z=1$, and extract the other
exponents using finite-size scaling of several physical observables.

\subsection{Paper Outline}

In Sec.~\ref{trial&aspect} we review the PQMC method for quantum spin systems in the valence bond (VB) basis, to make clear the role of the
boundary conditions of the time evolution. We also introduce the four trial states used in the simulations and define the physical observables
and their corresponding PQMC estimators that we use to study the critical fluctuations. In particular, we discuss the estimator of the spin
stiffness, which so far has only been derived within $T>0$ QMC methods but for which we here present a simple generalization for PQMC
calculations. In Sec.~\ref{properties} we discuss the finite-size scaling ansatz within which we analyze our data. We then
determine the critical coupling of the bilayer model and extract its universal exponents at criticality. We also discuss atypical
and non-asymptotic properties that originate from the initial trial states and finite projecting time. We summarize the results
and further discuss them in Sec.~\ref{conclusion}.

\section{Valence-Bond Projector Method}
\label{trial&aspect}

In a PQMC simulation, the ground state $|0\ra$ of a system is reached by projecting a trial singlet state $|\Psi(0) \ra$, as described
in Eq. (\ref{UPsi}). For a spin-isotropic, bipartite $S=1/2$ quantum spin systems, the sampling of $Z_\tau$ can be carried in the restricted
VB basis,\cite{liang,pqmcliang,pqmc} where the VBs (singlets) connect sites only on different sublattices. The arbitrary trial singlet
state $| \Psi(0) \ra$  is expressed in the VB basis as
\be
|\Psi(0) \ra=\sum_r w_r |V_r\ra,
\label{aps1}
\ee
where $|V_r\ra$ is a tiling of $N/2$ singlets $(a,b)=\uparrow_a\downarrow_b - \uparrow_b\downarrow_a$ on a lattice with $N$ sites, with
$a$ and $b$ referring to sites on sublattice A and B, respectively, and the coefficients $w_r$ are all positive. These conventions
correspond to Marshal's sign rule for the ground state of a bipartite systems.

\subsection{Sampling space}

An antiferromagnetic Heisenberg Hamiltonian can be written as a sum $-\sum_{ij}J_{ij}P_{ij}$ of singlet projectors,
\be
P_{ij}=  1/4-{\bf S}_i \cdot {\bf S}_j.
\label{pij}
\ee
When projecting with a high power $(-H)^n$ of the Hamiltonian, a PQMC configuration corresponds to a string of $n$ of these singlet projectors
acting on a component (a singlet tiling) of the VB trial state. Instead of the fixed power, one can also, as we will do here, use the Taylor
expansion of ${\rm e}^{-\tau H}$ and sample strings of a fluctuating number $n$ of operators. When a singlet projector acts on a VB state, either
the two sites $ij$ are connected by the same bond, in which case the state stays unchanged, or the sites belong to two different bonds that
become reconfigured so that one of the new bonds connects sites $i$ and $j$ and the second one connects the two sites that were previously
connected to $i$ and $j$. The latter process comes with a factor $1/2$ in the weight of a configuration, while the factor is unity in the former
case. One can formulate a PQMC algorithm based on these simple rules purely in the VB basis,\cite{pqmc} but the sampling is rather slow compared to
state-of-the art $T>0$ methods.

By sampling the spin configurations corresponding to a given singlet tiling of the trial state, one can formulate a PQMC algorithm that is very
similar to the $T>0$ SSE method running at $\beta=2\tau$, including very efficient loop updates.\cite{pqmc1} Let us compare the two approaches.
In the SSE method, ${\rm e}^{-\beta H}$
is Taylor expanded up to all contributing orders $n$, with the maximum contributions of order $n = -\beta \langle H\rangle \propto \beta N$. The
expansion order $n$ is sampled in the simulation according to the total weight of contributions from that order. Often, for practical reasons,
a self-selected upper bound $n_{\rm max}$ is imposed, so that the sampling scheme can be formulated with a fixed number of operators
in the operator strings corresponding to the evolution of traced-over states $|\alpha\rangle$ in the chosen basis (normally the basis of $S^z_i$
spins). Then $n$ of these $n_{\rm max}$ operators are selected among the terms $H_b$ of the Hamiltonian and the remaining $n_{\rm max}-n$
entries are ``fill-in'' identity operators.

Essentially, going from the SSE to the PQMC method corresponds to opening up the periodic time boundaries arising from the trace
operation, formally replacing the trace $\sum_\alpha \langle \alpha|\cdots |\alpha\rangle$ by a sum
$\sum_{\alpha_L,\alpha_R} c_Lc_R\langle \alpha_L|\cdots |\alpha_R\rangle$ corresponding the projection out of the trial state. In the
valence bond basis, this results in ``sealing" open loop segments at the time boundaries with VBs.
Then, as in the SSE case, a space-time spin configuration can be fully decomposed into loops that can be flipped independently of each other.
One can show that all signs cancel out in the overall weights for the PQMC configurations, because of the bipartiteness of the system.

\begin{figure}
  \includegraphics[width=8.5cm]{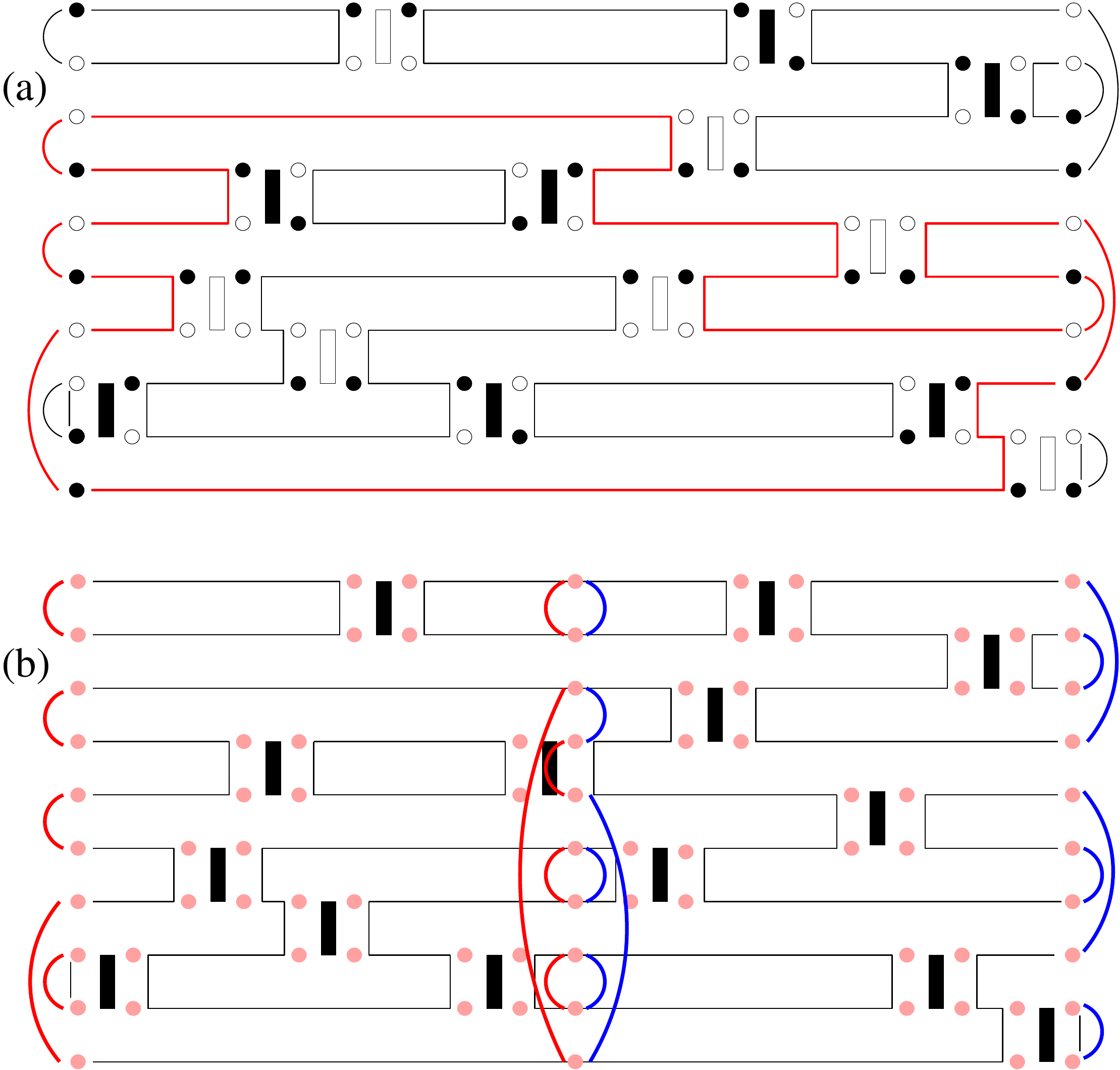}
  \caption{Two different representations used for the configuration space in PQMC simulation in the VB basis, exemplified with
    a Heisenberg chain of 10 sites. VBs of the sampled trial state are shown at the left and right edges. In (a), the linked vertex
    representation used for loop updates is shown, with solid and open circles indicating up and down spins. Diagonal and off-diagonal
    operators, $-(S^z_iS^z_j-1/4)$ and $-(S^+_iS^-_j+S^-_iS^+_j)$ are denoted by open and solid vertical bars, respectively. An example
    of a loop is shown in red; flipping this loop changes all up spins to down, and vice versa, thereby also changing the operator type
    (diagonal or off-diagonal) on some of the vertices. The number of operators $n$ and their locations on the lattice are updated
    separately in ``diagonal updates''. In (b), the pure VB representation used for collecting operator expectation values is shown
    for the same configuration. Here there is no distinction between diagonal and off diagonal operators and the bars indicate the
    full singlet projectors $P_{ij}$. Propagated left and right VB configurations are shown at the center; they make up the transition
    graph on which correlation functions are evaluated.}
  \label{pqmcfig}
\end{figure}

The spin and valence bond representations of a configuration are illustrated by an example in Fig.~\ref{pqmcfig}, including the concept
of loop updates in the spin representation, Fig.~\ref{pqmcfig}(a).
The VBs in the trial state can be updated by simple reconfigurations of pairs of bonds, to maintain the A-B connectivity with
anti-parallel spins on each bond, using the weights $w_r$ in standard Metropolis acceptance probability. One can also formulate loop
updates of the trial state,\cite{pqmc1} but this is more useful in variational calculations than in PQMC.

When evaluating operator expectation values, it is normally (depending on the type of operator considered) better to return to the VB
only basis, illustrated in Fig.~\ref{pqmcfig}(b), which corresponds to summing over all spin configurations that are compatible with the VBs
in the trial state and the lattice locations of the operators in the string---here also all the operators turn into the full singlet
projectors $P_{ij}$, instead of their individual diagonal and off-diagonal terms in the spin basis.

In principle, we do not have to sample VBs of the trial state; we can just use a fixed configuration of the VBs and that will still have
overlap with the ground state and converge to the same. However, one can also construct good, practically workable variational states and
this can improve the convergence properties.\cite{pqmc1}
Even without optimizing, one can write down simple translationally invariant states to further
restrict the simulations to total momentum $k=0$. In $T>0$ simulations, states with all ${\bf k}$ and all total spin values $S$ are included,
and to reach the ground state $T$ has to be well below the smallest gap from the $k=0,S=0$ ground state. PQMC simulations are restricted
by construction to  $S=0$ and $k=0$, and, thus, most of the low-energy states are excluded from the outset and do not have to be projected away.
Here our purpose is not to reach the ground state perfectly, and we will test the typicality hypothesis both with $k=0$ trial states and with
simple ``frozen'' VB states that do not conserve ${\bf k}$.

\subsection{Different Trial States}

The amplitude-product states proposed by Liang {\it et al.} \cite{pqmcliang,liang} are good $k=0$ variational states to describe
N\'eel ordered, critical, and quantum paramagnetic systems. The wave-function coefficients are of the form
\be
w_r=\prod_{i=1}^{N/2} h({\bf r}_i),
\label{aps2}
\ee
where ${\bf r}_i$ denotes the ``shape'' of the $i$-th singlet (the lengths of the VB in all lattice directions)
and $h({\bf r}_i)>0$ should respects all lattice symmetries of a translationally invariant system. A commonly used form
is $h({\bf r})=|{\bf r}|^{-\alpha}$ with $|{\bf r}|$ the length of the VB and $\alpha>0$ an exponent that can be optimized;
for example, for the 2D Heisenberg model the best choise is $\alpha=3$.\cite{lou07} Variational optimization of the amplitudes
can give extremely good energies---probably the best variational energies ever achieved for Heisenberg models.\cite{lin12,mlnote}
Here again we primarily want to compare different trial states, and we define $|\Psi_1\ra$ with $\alpha=3$ and $|\Psi_2\ra$ with $\alpha=6$.

As the third trial state we choose a single VB configuration,
\be
|\Psi_3\ra=\prod^{L^2}_{i=1}  (i_1,i_2),
\label{vs1}
\ee
which is a product state of singlets $(i_1,i_2)$ on the vertical bonds connecting two adjacent sites $i_1$ and $i_2$ in different layers.
It is the asymptotic ground state of the Hamiltonian (\ref{H_blyer}) in the limit $J_2 \to \infty$; thus the time evolution can be understood
as an imaginary-time quench in the couplings of the system from $g=\infty$ to $g$ close to $g_c$ (when studying criticality), followed
by a waiting time $\tau$.

The last trial state  is again a simple VB configuration, but, unlike $|\Psi_3\ra$, it breaks the translational symmetry of the system.
We choose a columnar arrangement of the VBs in  each layer;
\be
|\Psi_4\ra=\prod_{a=1,2}\prod_{x'_{ai}} (i_a,i_a+\hat{x}),
\label{vs2}
\ee
where $i_a+\hat{x}$ stands for the site shifted from $i_a$ by one lattice spacing along the $x$ direction and $x'_{ai}$ denotes
a site $i$ of layer $a$ whose $x$ coordinate is odd.

The energy expectation values $E=\langle H\rangle/N$ of the above trial states can be evaluated by sampling the VBs of $|\Psi_1\rangle$
and $|\Psi_2\rangle$, and for $|\Psi_3\rangle$ and $|\Psi_4\rangle$ exact calculations are trivial. For $L=16$, as an example,
the results at the estimated critical point
$g=2.5222$ (see further below) are $E_1=-1.841(1)$, $E_2=-1.829(1)$, $E_3 =-1.7611$, and $E_4=-1.190275$. The unbiased, sufficiently projected
energy is $E=-1.94225(2)$. Obviously, the last two trial states are far away from good variational states describing the criticality of the current
model. Even $|\Psi_1\rangle$ and $|\Psi_2\rangle$ are not very good variational states, and much better ones can in principle be obtained by
optimizing the bond amplitudes. However, our purpose here is not to optimize the states and the simulations, but to demonstrate that typical
critical fluctuations emerge out of arbitrary trial states with projection time $\tau \propto L$. For this purpose the above range of
trial states will suffice.

\subsection{Physical Observables and PQMC Estimators}

The staggered magnetization, the order parameter, is defined as
\be
{\bf m}=\frac{1}{N}\sum_{i=1}^N (-1)^{x_i+y_i+z_i}{\bf S}_i,
\ee
where $(x_i, y_i, z_i)$ are the integer coordinates of the spin at site $i$ of the bilayer with $N=2L^2$ sites.
The squared magnetization $\langle {\bf m}^2\rangle$
can be efficiently estimated in PQMC simulations using the sum of squared loop lengths in the transition graph obtained by superimposing
the sampled ``left'' and ``right'' projected VB configurations,\cite{vbsv} as illustrated in Fig.~\ref{pqmcfig}(b).

The Binder ratio \cite{Binder} is defined as
\be
Q=\frac{\la {\bf m}^4\ra}{\la {\bf m}^2\ra^2},
\ee
where ${\bf m}^4$ can also be calculated according to the loop structure of the transposition graphs.\cite{vbsv}

The spin stiffness $\rho_s$ characterizes the tendency of ordered spins to adapt in response to a twist imposed on the spins in an
ordered state in a direction perpendicular to the ordering vector. In the common QMC simulations at finite temperature, e.g., with the SSE
method \cite{qspin} or path integrals,\cite{Pollock} there is a very convenient estimator based on fluctuations of the winding number
characterizing the topology of the spin world lines propagated around the space-time periodic system. The stiffness along the $\alpha$ lattice
direction is given by
\begin{equation}
\rho_s=\frac{\langle W_{\alpha}^2\rangle}{L^{d-2}\beta},
\label{rsse}
\end{equation}
with the size normalized winding number defined in the SSE method as
\begin{equation}
 W_{\alpha}=\frac{1}{L}(N_\alpha^+-N_\alpha^-).
\label{winding}
\end{equation}
Here $N_\alpha^+$ and $N_\alpha^-$ denote the total number of off-diagonal operators transporting spin in the positive and negative
$\alpha$ direction, respectively. So far, the spin stiffness has not been considered in PQMC simulations, as far as we are aware,
likely because the winding number is not a well defined conserved topological number in this case. However, it is still possible
to proceed with an unbiased generalization of the above winding number estimator, as we describe next.

The winding number formula (\ref{winding}) is a consequence of the periodic boundaries in both the spatial and time directions at $T>0$.
The "cutting open" of the time boundaries in the PQMC method makes this estimator fail at first sight. However, since the current fluctuations
within some local time segment $\Delta\tau$ should be independent of $\beta$ for large $\beta$, and these fluctuations are what gives rise
to the winding numbers, it is clear that the conservation of the winding number is not very consequential in Eqs.~(\ref{rsse}) and
(\ref{winding}), but is just a byproduct of the
periodic time boundaries in combination with the conserved magnetization of the system. Thus, it should be possible to generalize
the formula by considering generalized, non-integer  winding numbers over a sufficiently large time interval $\Delta\tau \le 2\tau$,
where we recall that the total length of the system in imaginary time is $2\tau$. Since the time dimension is not uniform, due to the
open boundaries, one can also presume that convergence to the correct value when $2\tau, \Delta\tau \to \infty$ will be faster if
$\Delta\tau < 2\tau$ and the interval is time-centered.

As a reasonable choice satisfying the above requirements, we take the centered interval with $\Delta\tau = \tau$ and calculate
the spin stiffness of a $d$-dimensional system according to
\begin{equation}
\rho_s  =\frac{ \langle (\tilde N^+_\alpha-\tilde N^-_\alpha)^2\rangle}{ \tau L^d}
\label{rhosPQMC1}
\end{equation}
where $\tilde N^+_\alpha$ and $\tilde N^-_\alpha$ denote the total number of operators transporting spin in the positive and negative
$\alpha$ direction by the middle part of the operator string. We have used the fact that the total length is linearly proportion to
$2\tau$, which is $\beta$ in a corresponding SSE simulation.

\begin{figure}[t]
 \includegraphics[width=7cm]{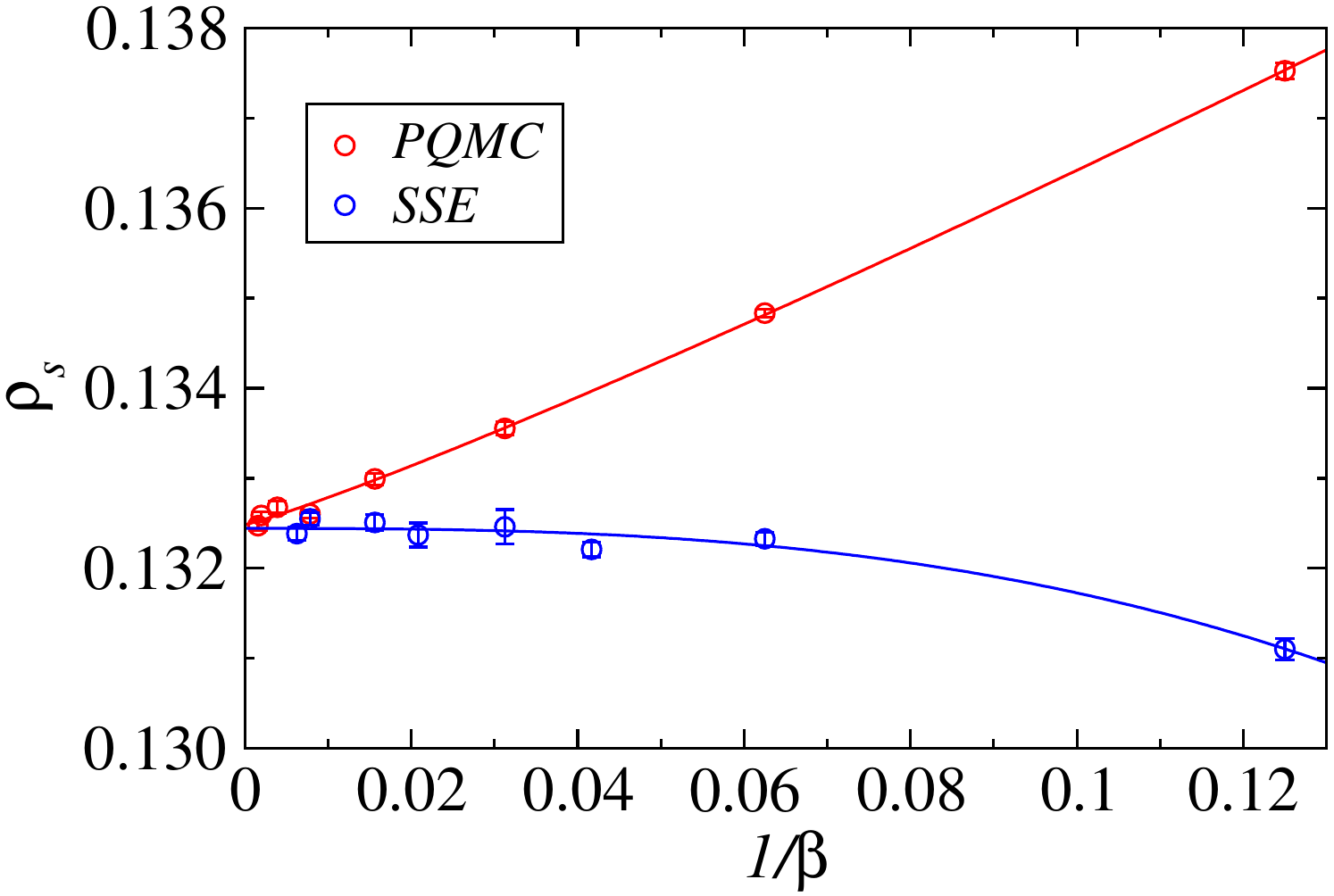}
 \caption{Spin stiffness of the 2D Heisenberg model on the square lattice of linear size $L=16$ as a function of inverse temperature
   $\beta$ in SSE simulations or the projection time $2\tau=\beta$ in PQMC simulations. The results were obtained using the standard
   winding number fluctuations according to Eq.~(\ref{rsse}) and the corresponding generalized spin current formula (\ref{rhosPQMC1}),
   in the SSE and PQMC simulations, respectively.}
 \label{spinstiffness}
\end{figure}

To validate the above formula, we simulate the square lattice Heisenberg model with system size $L=16$ using both the SSE
and PQMC methods. Results obtained by the two methods are shown versus $\beta$ and $2\tau$ in Fig. \ref{spinstiffness}). As the
ground state is approached for large $\beta$ and $\tau$, we can see convergence to the same value. We find $\rho_s=0.13239(2)$ from
winding number fluctuations in the SSE simulations and  $0.1324(1)$ from the generalized spin current definition in Eq.~(\ref{rhosPQMC1})
in the PQMC simulations.
The two estimates agree perfectly. In the case of SSE, we expect asymptotic exponentially fast convergence (though the above result
was obtained by extrapolation using a polynomial, which should be fine at the level of the statistical error bars), reflecting the
finite-size gap in the spectrum, while in the case of the PQMC calculations with the estimator (\ref{rhosPQMC1}) the convergence appears
to be linear in $1/\tau$. We do not currently have an understanding of this behavior, though clearly it must be related to the fact
that the $\Delta\tau$ region over which the current fluctuations are summed have open boundaries and one may expect a correction
proportional to the inverse of the length of the boundary. Thus, SSE may still be the preferable way to compute $\rho_s$, though certainly
this test (and others) shows that this important physical quantity can also be reliably obtained  in PQMC calculations.

\section{critical properties}
\label{properties}

\subsection{Finite-size Scaling Ansatz}

For a singular quantity $A \propto \delta^{\kappa}$ in the thermodynamic limit, according to the standard finite-size scaling theory,\cite{fss}
scaling of the following form is expected close to the critical point $g_c$:
\be
A(g, L)= L^{-\kappa/\nu} f(\delta L^{1/\nu},L^{-\omega}),
\label{fssA}
\ee
where $\delta=(g-g_c)/g_c$, the exponent $\kappa$ depends on the quantity in question, and both $\kappa$ and $\nu$ are tied to the
universality class of the transition. We have included only the most important scaling correction, with the associated exponent $\omega>0$.
Exactly at $g_c$ ($\delta=0$), and neglecting the scaling correction for now, the form reduces to
\be
A(g_c,L) \propto L^{-\kappa/\nu},
\label{fssA@qc}
\ee
given that the non-singular scaling function $f$ approaches a constant when $\delta \to 0$.

To locate the critical point, we can treat the scaled $A(g,L)L^{\kappa/\nu}$ as a dimensionless quantity $\tilde{A}(g, L)$. By  Talyor
expanding the scaling function in (\ref{fssA}) and keeping the correction, we have
\be
\tilde{A}(g,L)=\tilde{A}_c+a_1\delta L^{\frac{1}{\nu}}+a_2\delta^2L^{\frac{2}{\nu}}+b_1L^{-\omega}
+c_1\delta L^{\frac{1}{\nu}-\omega}+\cdots
\ee
where $\tilde{A}_c, a_1, a_2, b_1, c_1$ are unknown, non-universal constants.
This implies that curves $\tilde{A} (g, L_1) $ and $\tilde{A}(g,L_2) $ versus $g$ cross each other at some
$g=g^*(L_1, L_2)$. We will take $L_1=L$ and $L_2=2 L$, for which the crossing point $g^*(L)$ approaches $g_c$ as \cite{luck85}
\be
g^*(L) - g_c \propto L^{-(1/\nu +\omega)}.
\label{gc}
\ee
The critical point $g_c$ can thus be extrapolated.
The critical value of $\tilde{A}_c$ may also be universal and is therefore interesting. It can be extracted by calculating
the  quantity $\tilde{A}^*(L)\equiv \tilde{A}(g^*, L)$ at the crossing point $g^*(L)$,  which approaches its limit ${\tilde A}_c$ in the following way
\be
{\tilde A}^*(L) -\tilde{A}_c \propto L^{-\omega}.
\label{Ac}
\ee
In principle both $\omega$ and $\nu$ can be extracted from Eqs.~(\ref{gc}) and (\ref{Ac}), though in practice the neglected higher-order
corrections often distort the values significantly. One can instead extract $1/\nu$ from the slope of ${\tilde A}^*(L)$ at the
crossing point, as described in many papers (including a recent systematic study in the Supplemental Material of Ref.~\cite{shao2}).

Alternatively the correlation exponent $\nu$ can also be estimated by staying at the size-extrapolated critical point, if this point has
been located to sufficient precision. We will use this approach here. First, we calculate the derivative $s(g_c, L)$ of $\tilde{A}$ to $g$
at the estimated critical point $g_c$. This is done by fitting a polynomial $f(g)=ag^2+b g +c $, with three unknown constants $a, b, c$, to
six values of $\tilde{A}(g, L)$ near $g_c$. Error bars can be estimated by Gaussian noise propagation. Then, according to the following
scaling formula,
\be
s(g_c, L)=\frac{{\rm d} \tilde{A}(g,L)}{d g}|_{g_c}=a_1 L^{\frac{1}{\nu}}(1+b L^{-\omega}),
\label{snu}
\ee
we find $\nu$ by using nonlinear fits to the slopes. We will here exclude small system sizes so that the correction can be safely
neglected, given that the correction exponent is relatively large; $\omega \approx 0.78$.\cite{Vicari}

The Binder ratio $Q$ is dimensionless, which means $\kappa=0$, while for the spin stiffness $\rho_s$, $\kappa=(d+z-2) \nu$, or $\kappa=\nu$
in the present case where $d=2, z=1$.\cite{Fisher} Therefore $Q$ and $\rho_s L$ are useful observables for locating the critical point and
estimating the correlation exponent $\nu$.

For the order parameter ${\bf m}$, $\kappa$ is the exponent $\beta$, which leads to the scaling behavior of the squared staggered magnetization,
\be
\la {\bf m}^2\ra \propto L^{-2\beta/\nu}(1+c L^{-\omega}),
\label{m2L}
\ee
at the critical point, with $c$ a constant. According to the scaling relation $2\beta/\nu = 1+ \eta$, we can estimate $\eta$ from
the size dependence, where, again, we typically do not include the correction term.

\subsection{The Critical Point}

We next study the typical behavior of the critical fluctuations, by performing PQMC simulations of the bilayer Heisenberg model with the four
different trial states defined above and with different prefactors $a$ in $\tau = aL$ and using $L$ up to 128. We typically used $10^5$ MC steps
to equilibrate the system and $10^6$ for collecting data for the physical quantities of interest. To project out the ground state fully,
$\tau$ needs to satisfy $\tau \gg 1/\Delta$, with $\Delta$ the gap between the ground state and the first excited state ``seen'' in the
calculations, which in the VB basis is the second singlet state. If we can find good scaling properties even significantly away from
this limit, it means that a band of low-lying singlets also share the same critical fluctuations as the ground state.

\begin{figure}[t]
\includegraphics[width=7.5cm]{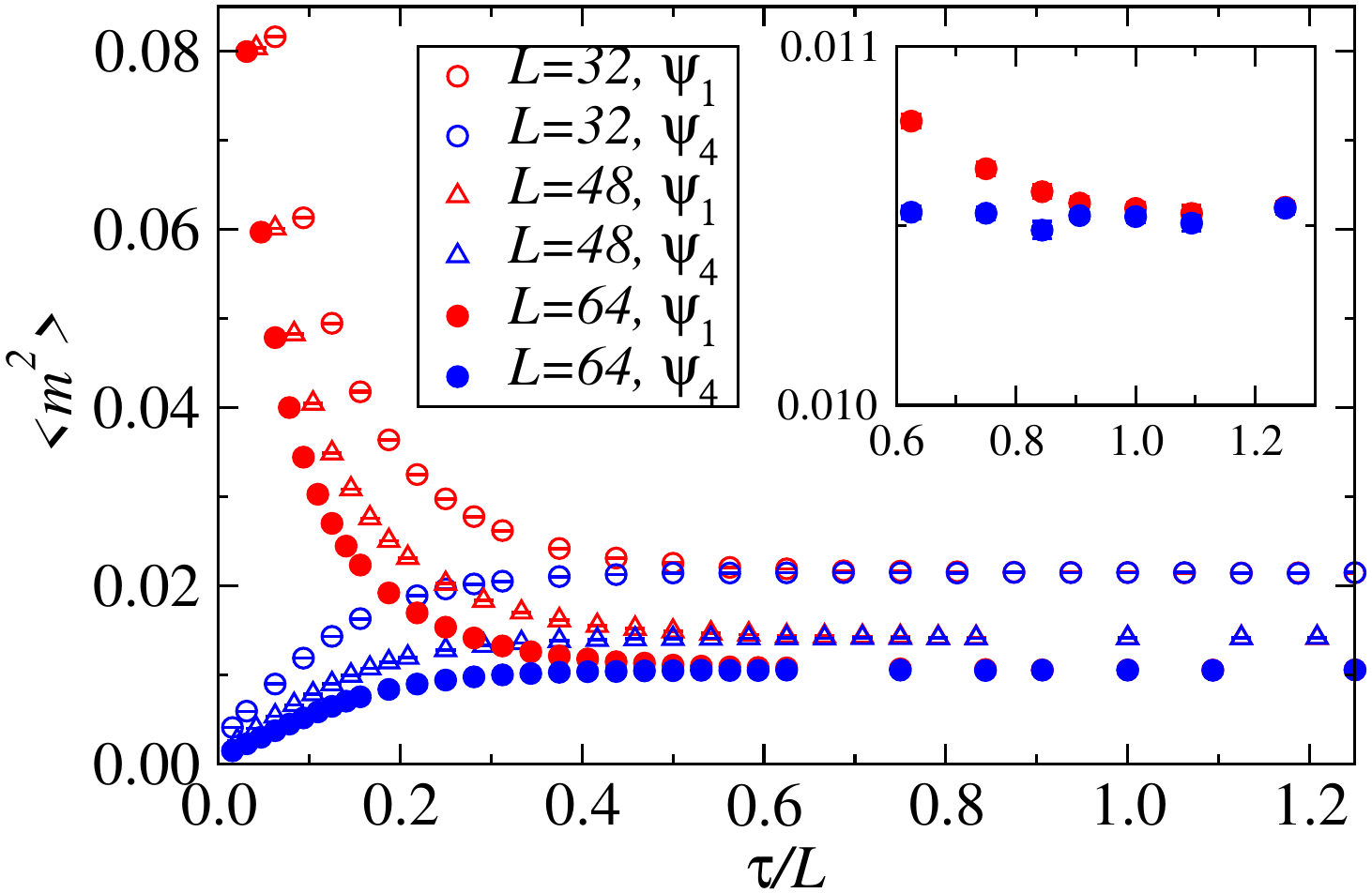}
\caption{The squared sublattice magnetization obtained in critical bilayer PQMC simulations with system sizes $L=32,44$, and $64$,
with the trial states $|\Psi_1\ra$ and $|\Psi_4\ra$ (representing the best and worst state as judged by the variational energy). The
inset shows the $L=64$ results close to $\tau=L$ on a more detailed scale.}
\label{mtau}
\end{figure}

In Fig.~\ref{mtau}, we first show results for the squared order parameter evaluated at the critical point, estimated below to
be $g_c \approx 2.5222$, as a function of the projection time for three system sizes and two trial states. On the scale used in the
figure, $\tau=L$ gives results almost indistinguishable from the ground state---a close examination (inset of the figure) reveals
that there are still some statistically significant differences. With the worst of the trial states $|\Psi_4\rangle$, the results have visibly
not converged for $\tau=L/2$, and at $\tau=L/4$ and smaller both trial states give results clearly different from the ground state.
Note that one $|\Psi_1\rangle$ has strong N\'eel that is decays away with increasing $\tau$, while $|\Psi_4\rangle$ has no long-range
correlations at all; thus the critical correlations are gradually emergent with increasing $\tau$. Thus, we have a range of different
trial states and it is interesting to see if the critical correlations can emerge universally even for small factors $a$ in
$\tau=aL$, where the behaviors in Fig.~\ref{mtau} look completely different for the two trial states.

\begin{figure}[t]
\includegraphics[width=7cm]{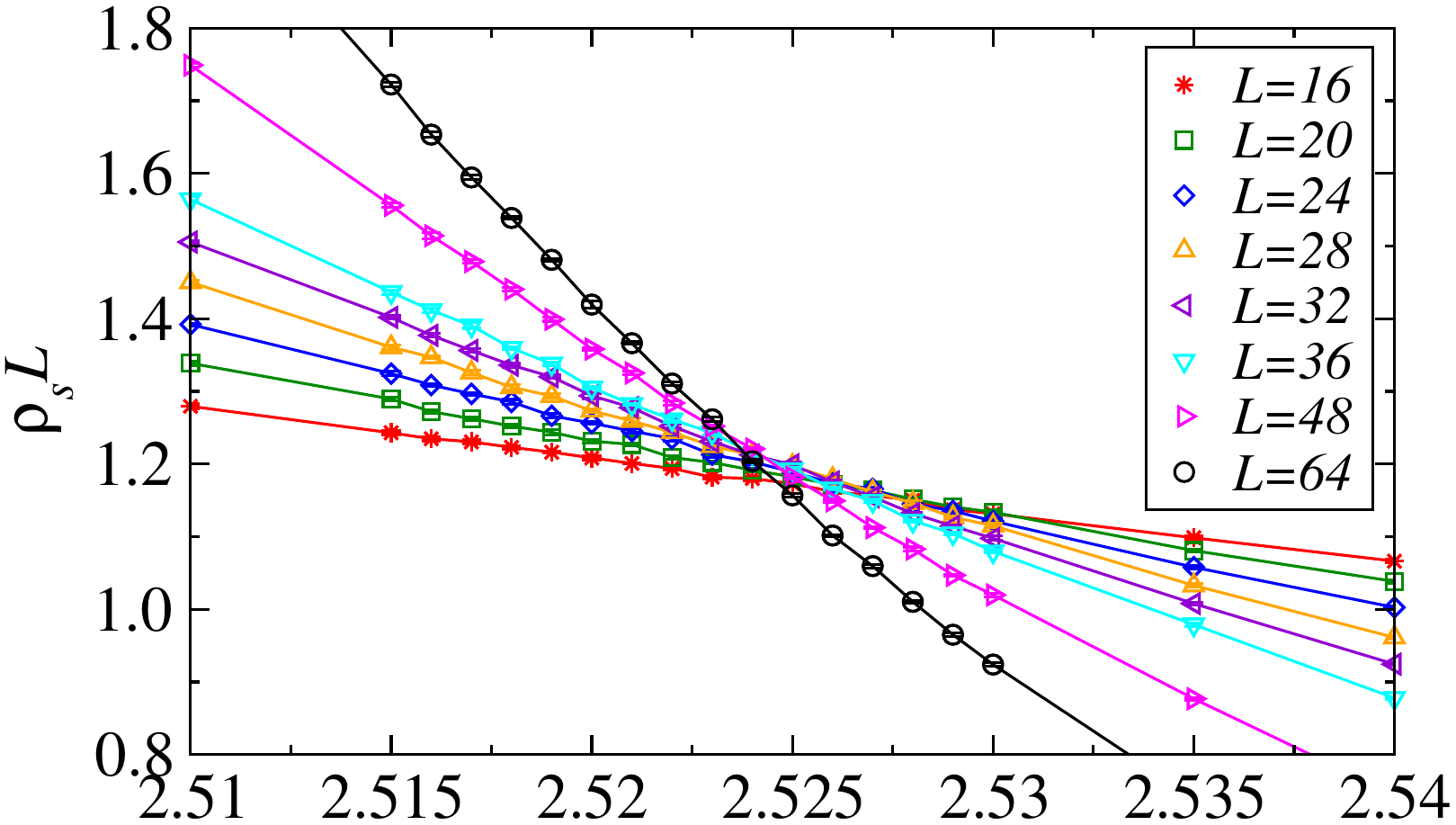}
\includegraphics[width=7cm]{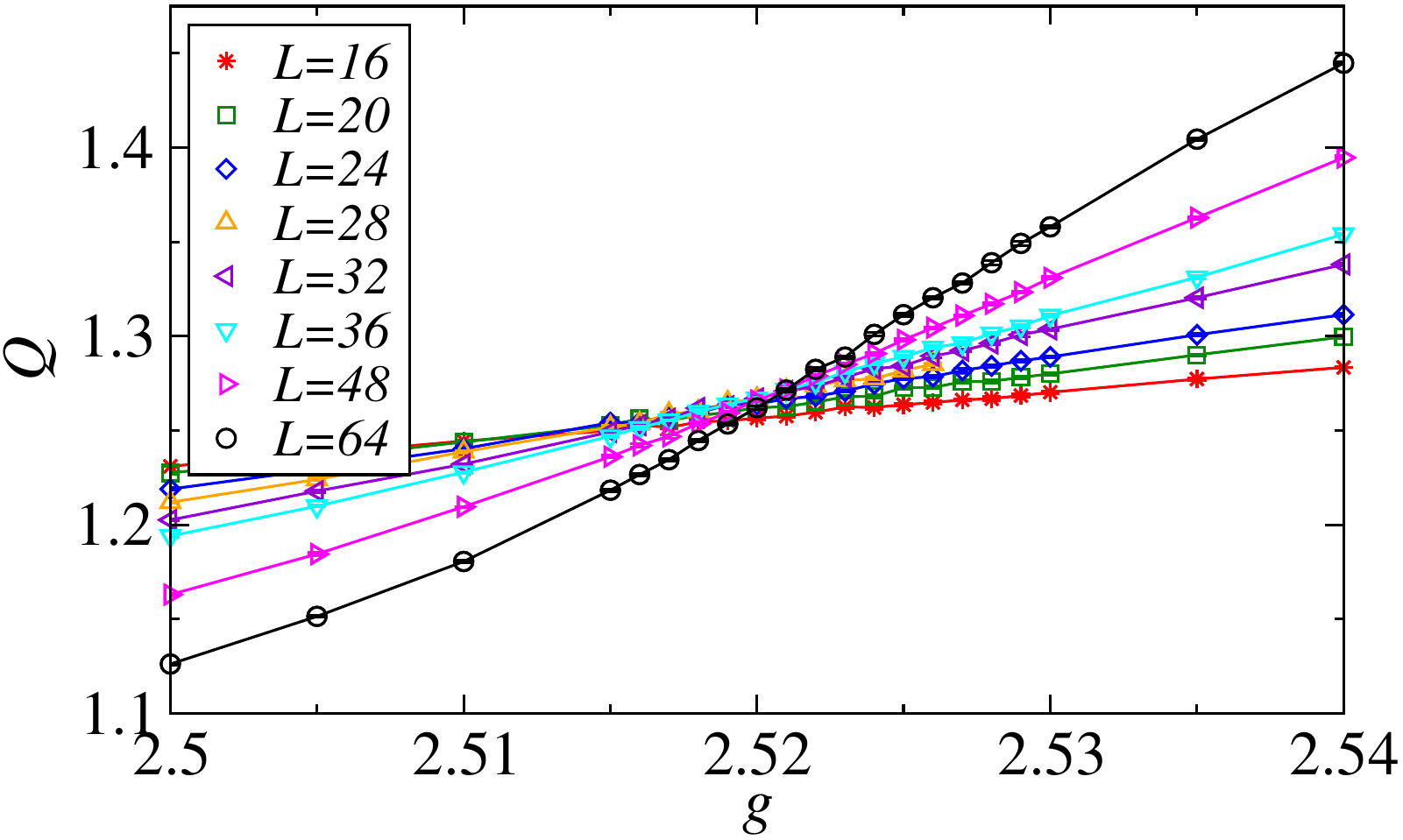}
\caption{The scaled spin stiffness $\rho_s L$ (a) and the Binder ratio $Q$ (b) of the bilayer Heisenberg model with several $L$
graphed vs the coupling ratio $g$. The PQMC calculations were carried out using $\tau=L$ ($a=1$).}
\label{stfdouble}
\end{figure}

\begin{figure}[t]
\includegraphics[width=7cm]{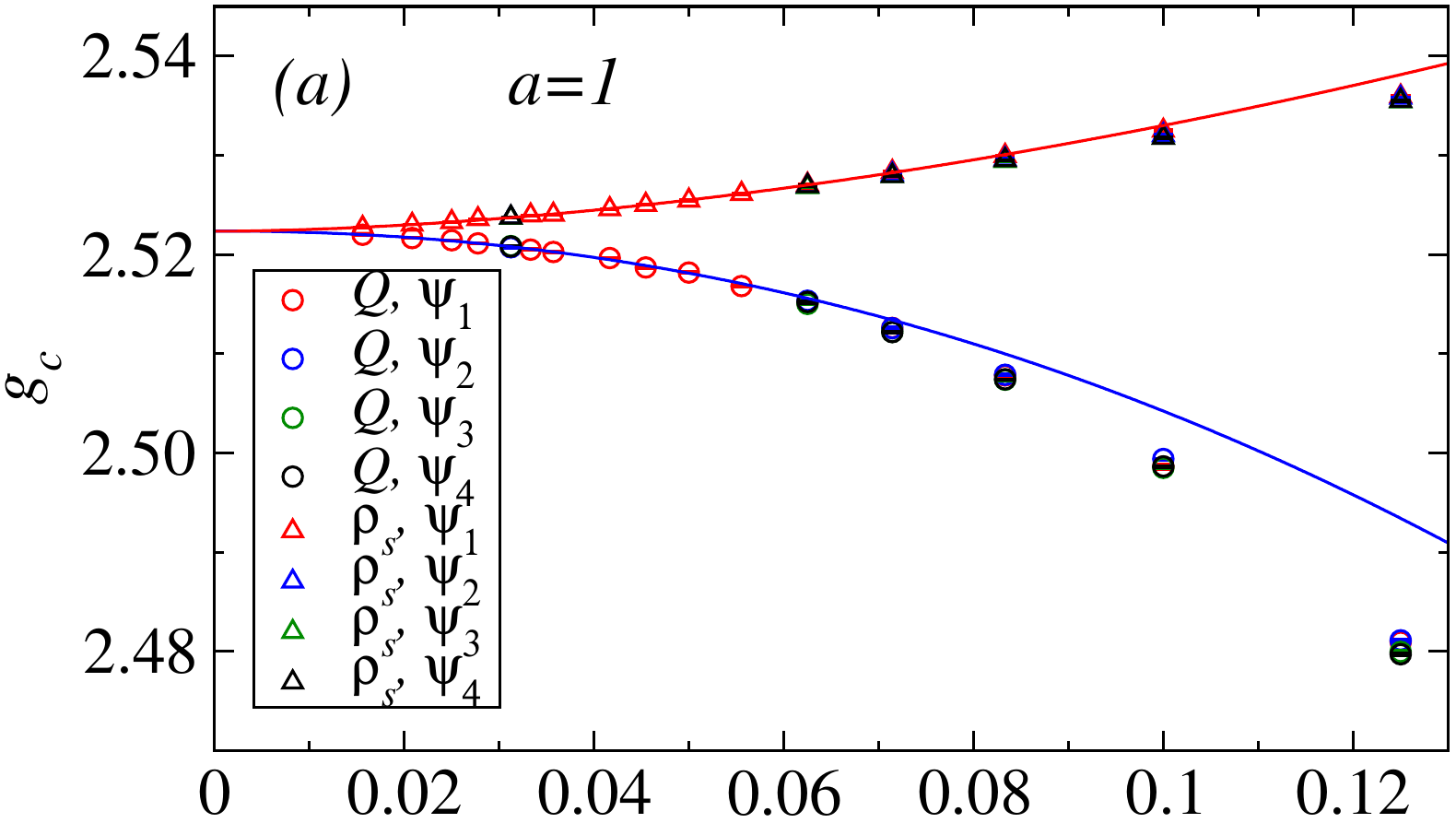}
\vskip2mm
\includegraphics[width=7cm]{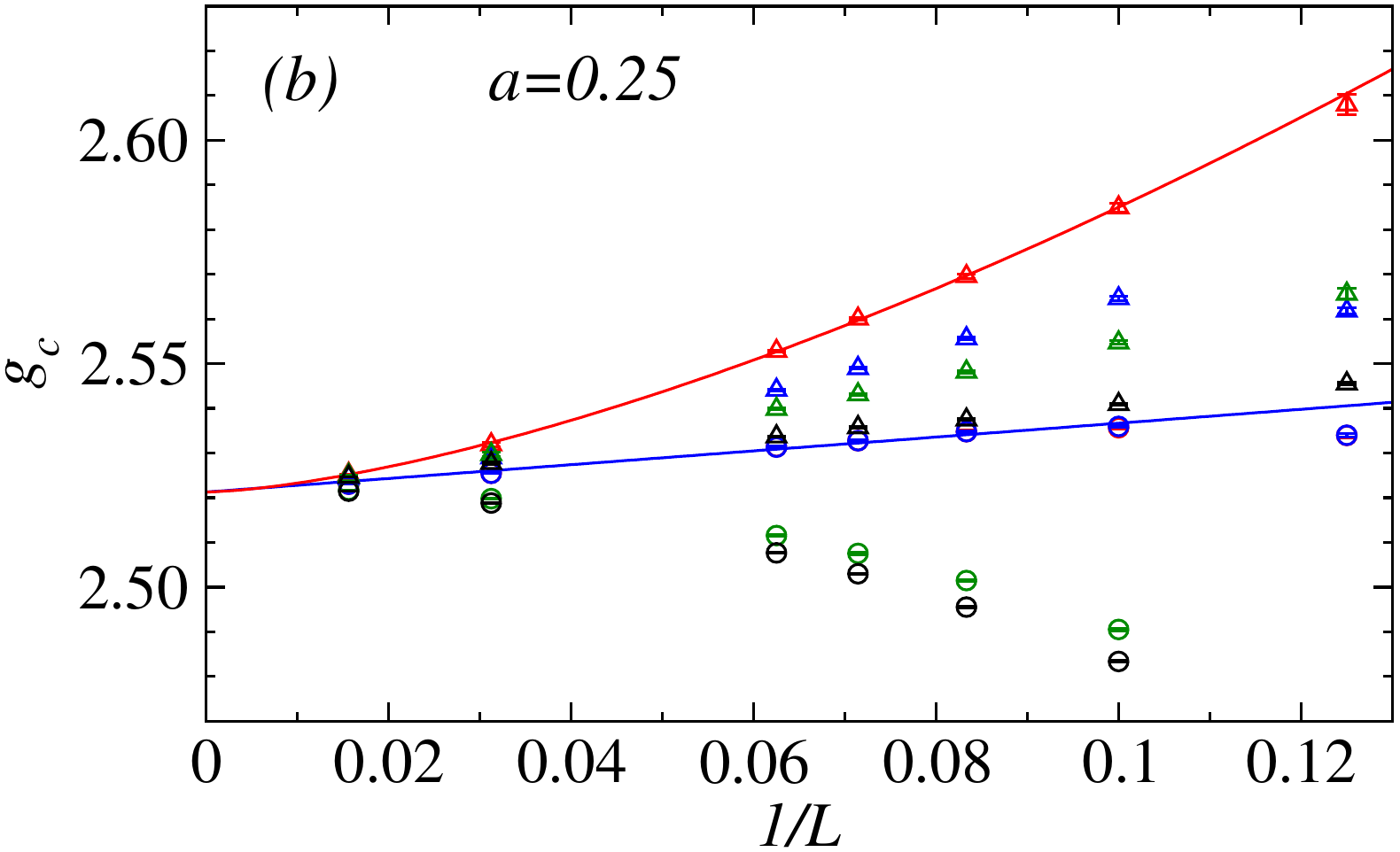}
\caption{Scaling of crossing points $g^*(L)$ of the size-scaled spin stiffness $\rho_s L$ (triangles) and the Binder ratio $Q$ (circles).
  The results were obtained from four different trial states, as indicated by the legends, and two different projection time factors
  $a$ were in the PQMC simulations; $a=1$ in (a) and $a=0.25$ in (b). The largest system size pair is $(L,2L)=(64,128)$. The solid curves
  are fits to $g^*$ from $\rho_s L$ crossings (red) and $Q$ (blue) obtained from states projected out of the best trial state, $|\Psi_1\ra$.}
\label{crossing}
\end{figure}

To analyze the critical point, we begin by considering the time regime
where we have almost reached the ground state, using the best trial state in the variational sense,
$|\Psi_1\ra$, and the projection time set to $\tau=L$, i.e., the factor $a=1$. The scaled spin
stiffness $\rho_s L$ and the Binder ratio $Q$ are shown versus the coupling ratio $g$ for various system sizes in Fig. \ref{stfdouble}.
The results do not differ appreciably from SSE results obtained at very low temperatures,\cite{dbAFH} indicating that the projection time
here brings us almost to the ground state. We find crossing points between results for system sizes $L$ and $2L$ using polynomials fitted
to the data points. These crossing points extracted for the two different quantities for a large number of size pairs are shown in
Fig.~\ref{crossing}(a). The drifts of the $g^*$ values obtained from both quantities are monotonic in $L$, and both of them converge
to a common critical point $g_c$ rapidly for large $L$.  All $L\ge 16$ points are consistent with the expected power-law, Eq.~(\ref{gc}).

A nonlinear fit of $g^*(L)$ from the $\rho_s L$ crossings yields $g_c=2.52222(4)$ and $1/\nu+\omega=1.65(3)$, with a reasonable reduced
goodness-of-fit value $\chi^2=1.2$. Here the result for the exponent combination $1/\nu+\omega$ is not very close to the expected O(3)
value, $1/\nu+\omega \approx 2.2$, likely reflecting the role of remaining higher-order corrections. Such still not size-converged
``effective exponents'' are known to not significantly effect the extrapolated critical point value.\cite{shao2} A similar fit of $g^*$ of
$Q$ gives $g_c=2.52224(6)$ and $1/\nu+\omega=2.3(1)$, with reduced $\chi^2=1.6$. Both estimates of critical point agree well with
earlier estimate $g_c=2.5220(1)$ obtained by using SSE QMC \cite{dbAFH}, in which the ground-state properties were obtained by
the $\beta$ doubling approach, but the statistical error is significantly reduced.

We next consider results obtained with the other trial states: $|\Psi_2\ra, |\Psi_3\ra$, and $|\Psi_4\ra$. The projection lengths
are first all set as $\tau=L$. The $(L, 2L)$ crossing points $g^*$ from $\rho_s L$ and $Q$ are both shown in Fig.~\ref{crossing}(a)
together with the previous results based on $|\Psi_1\ra$. We see only small differences between the results from the
different trial states and, not surprisingly, all these crossings converge to the common critical coupling $g_c$. Though it is not apparent
from the figure, somewhat larger system sizes are needed to fit the $|\Psi_{t>1}\ra$ results to power-law forms than what is the case
with $|\Psi_1\ra$. The latter state also is the best state in the sense of the variational energy. The results of all the fits are listed
in Tab.~\ref{table0}, including the smallest size used in the fit.

Since the above $a=1$ simulations deliver results quite close to the ground state for all the trial states, we need to go to smaller
$a$ to investigate the emergence of critical typicality in greater detail. Crossing points obtained from the $\tau=L/4$ simulations are
shown in Fig.~\ref{crossing}(b). Here we can see very significant differences from the $\tau=1$ data, but all crossing points still flow
toward the same critical point. Results of extrapolations are summarised in Tab.~\ref{table0}.

\begin{table}[b]
\caption{Results of finite-size analysis with a single power-law correction to the critical point $g_c$ for different trial states
and different value of the projection factor $a$. The standard goodness-of-fit per degree of freedom is denoted as $\chi^2$.}
\begin{tabular}{|c|c|c|c|c|c|c|}
\hline
\ &\multicolumn{6}{|c|}{$a=1$} \\
\hline
 trial state& $g_c(\rho_s)$ & $\chi^2(\rho_s)$  &$L_{\rm min}$& $ g_c (Q)$&$\chi^2(Q)$ &$L_{min}$ \\
\hline
$\Psi_1$ & 2.52222(4)& 1.2  & 8  & 2.52224(6) & 1.6 &16\\
\hline
$\Psi_2$& 2.5221(1)   & 0.5  &12 &2.5220(4) & 0.6 &12\\
\hline
$\Psi_3$ & 2.5220(2)  & 1.0&12& 2.5219(4) & 1.2 &12 \\
\hline
$\Psi_4$ & 2.5220(8)  & 0.5 &12 &2.5218(5) & 0.6      &12\\
\hline
\ &\multicolumn{6}{|c|}{$a=0.5$}\\
\hline
\   & $g_c(\rho_s)$ &$\chi^2(\rho_s)$& &$g_c(Q)$&$\chi^2(Q)$& \\
\hline
$\Psi_1$ & 2.5215(4) & 0.3 & 12&2.5221(1) & 0.7 &12\\
\hline
$\Psi_2$ & 2.5218(3) & 1.9 & 12&2.5221(4) & 0.8&12\\
\hline
$\Psi_3$ & 2.5220(2) & 0.6 & 12&2.5221(1) & 1.1 &12\\
\hline
$\Psi_4$  & 2.5220(2) & 0.6&12& 2.5221(2) & 1.5&12\\
\hline
\ &\multicolumn{6}{|c|}{$a=0.25$}\\
\hline
\   & $g_c(\rho_s)$&$\chi^2(\rho_s)$&&$g_c(Q)$&$\chi^2(Q)$&\\
\hline
$\Psi_1$ & 2.5213(2) & 0.9 &12& 2.5213(4) & 1.7&12 \\
\hline
$\Psi_2$ & 2.5222(2) & 0.8 &12& 2.5222(1) & 1.4&12\\
\hline
$\Psi_3$ & 2.5220(8) & 1.4 & 12&2.5223(3) & 0.4&12 \\
\hline
$\Psi_4$  & 2.5224(5) & 2.3 &12 &2.5212(3) & 1.4&12\\
\hline
\end{tabular}
\label{table0}
\end{table}

\subsection{Critical Exponents}

We next validate that the states projected out from various trial states at $\tau \propto L$ display typical critical fluctuations
characterized by the correct critical O(3) exponents. We demonstrate that this universality emerges with increasing system size
in finite-size scaling for all the different trial states.

\begin{figure}[t]
\centering
\includegraphics[width=7cm]{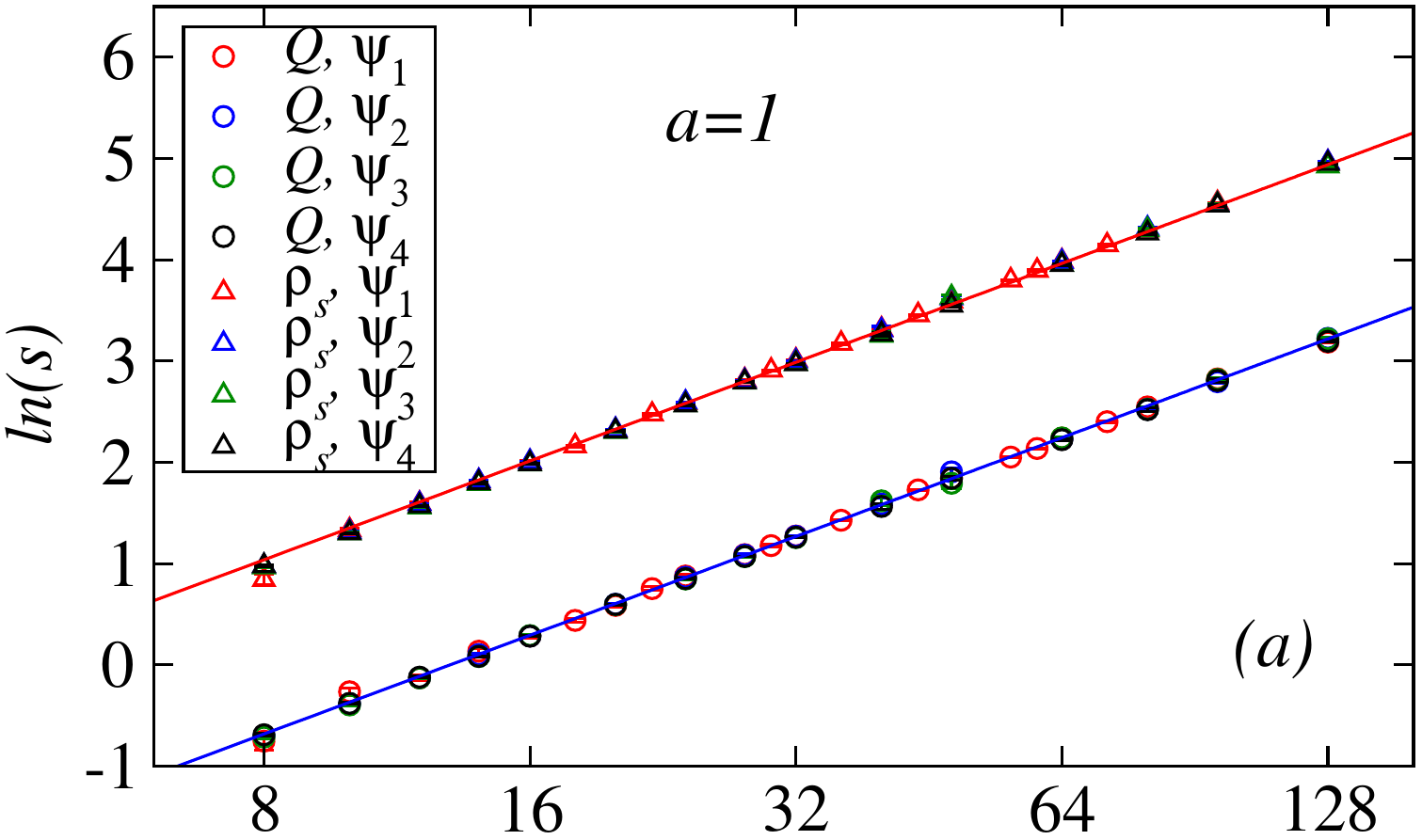}
\vskip2mm
\includegraphics[width=7cm]{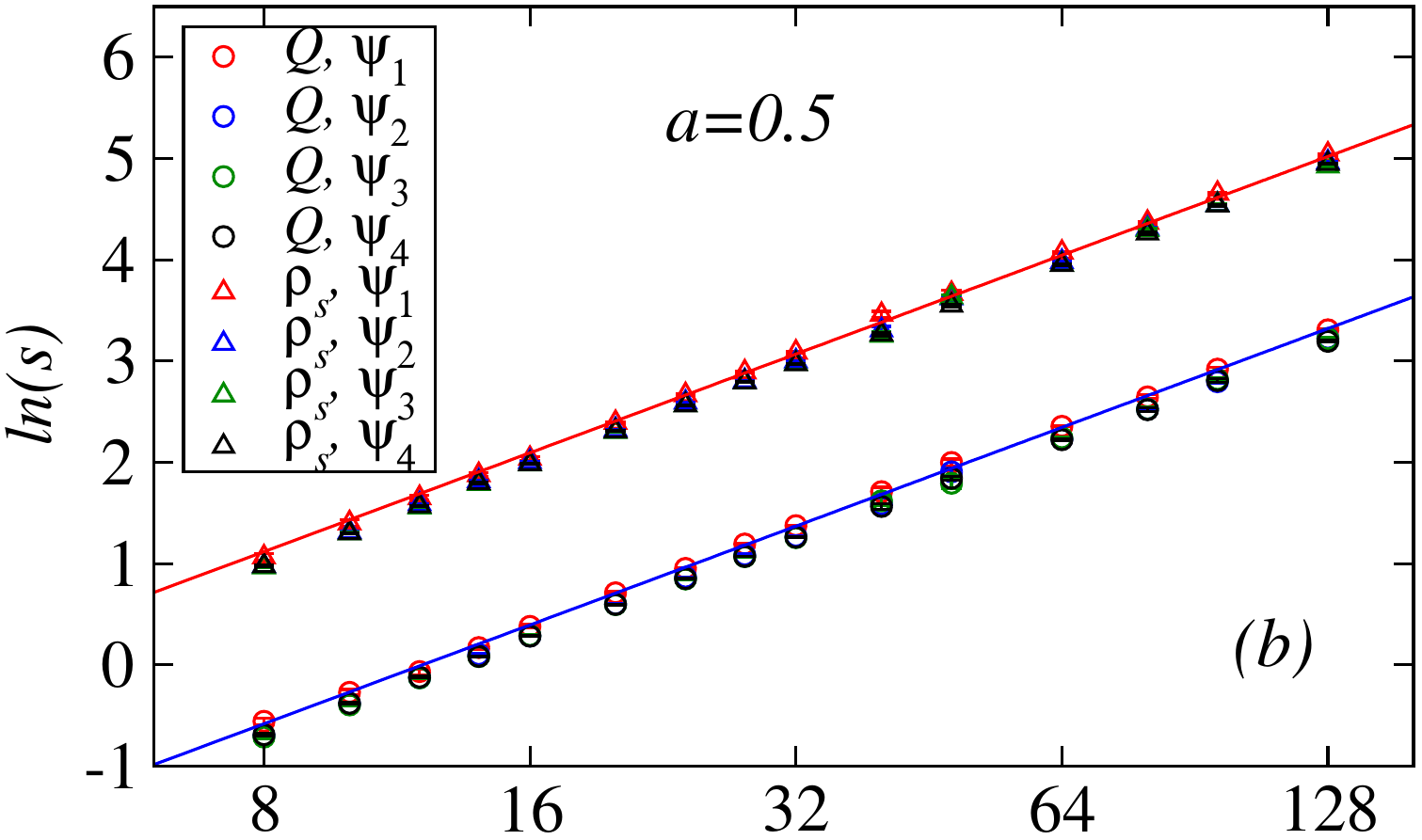}
\vskip2mm
\includegraphics[width=7cm]{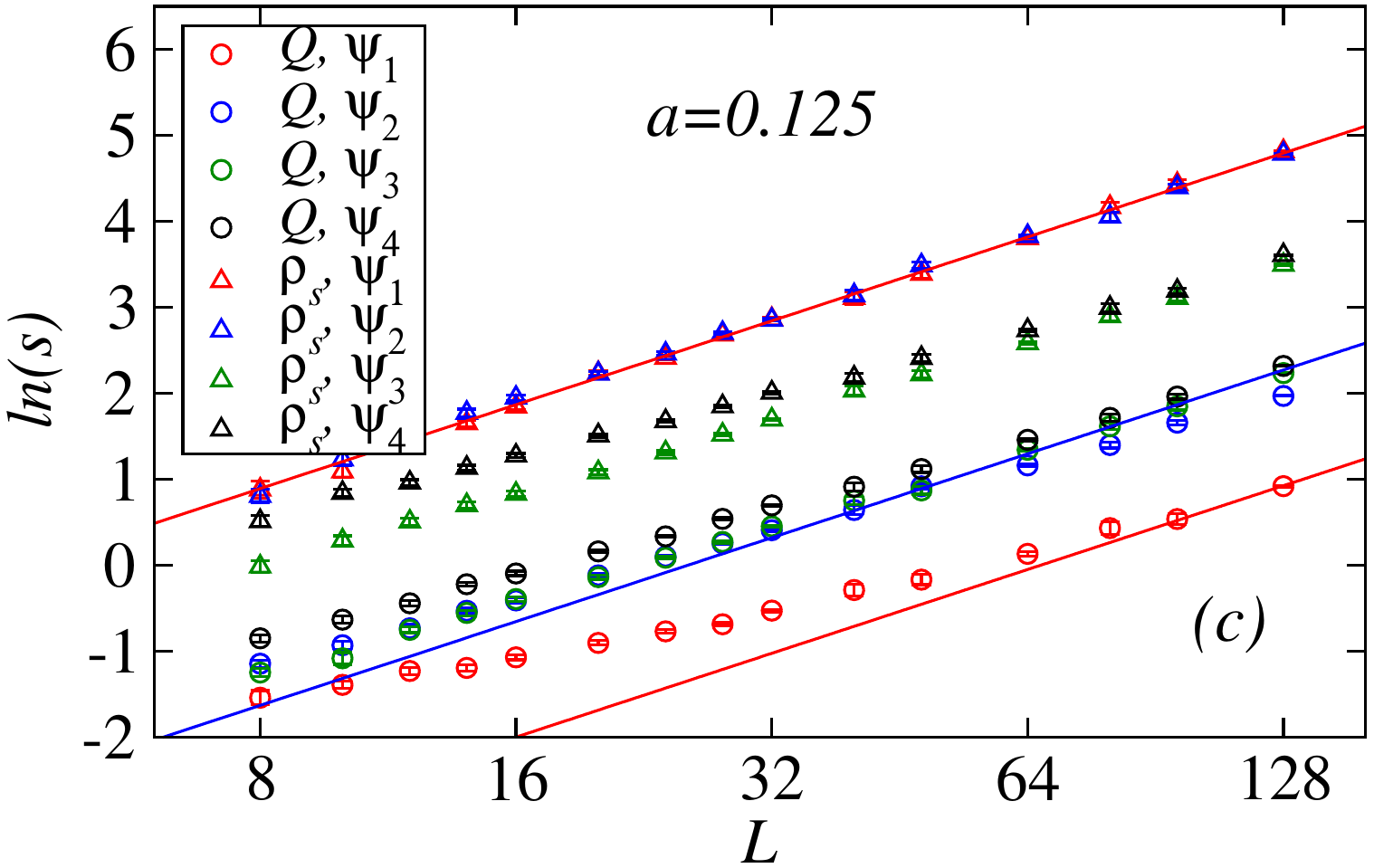}
\caption{The derivatives with respect to $g$ of $\rho_s L$ (triangles) and $Q$ (circles) at the best estimated critical point $g_c=2.5222$,
  graphed versus the system size on logarithmic scales. Three different projection times $\tau=aL$ were used; (a) $a=1$, (b) $a=0.5$,
  (c) $a=0.125$. Different graphing colors refer to results obtained with different trial states. The solid lines with slope corresponding
  to the known value of the exponent, $\nu=0.7112$, in Eq.~(\ref{snu}) are draw to show the expected large-$L$ behavior.}
\label{sL}
\end{figure}

First, we show that the correlation exponent $\nu$ can be extracted from the projected states. We extract the exponent according
to Eq.~(\ref{snu}) from the $g$ derivatives $s(g,L)$ of the curves $Q(g)$ at the estimated critical point, $g_c=2.5222$. The derivatives
are extracted from a quadratic polynomial fitted to a set of points for $g$ in the neighborhood of $g_c$. Results are shown in Fig.~\ref{sL}.

In the case $a=1$, for all four trial states and for data from both $\rho_sL$ and $Q$, the derivatives reproduce the expected power-law
behavior (\ref{snu}) as shown in Fig.~\ref{sL}(a). Here we do not include the correction term $L^{-\omega}$, where $\omega \approx 0.78$ is
expected for the universality class, as we find that the prefactor is small and good fits to just the leading power law $L^{1/\nu}$ can be achieved
if some of the smaller systems are excluded. Thus, we extract the correlation exponent $\nu$ from the data for each trial state by fitting
to Eq.~(\ref{snu}), starting from system sizes $L_{\rm min}$ sufficiently large for the quality of the fit to be acceptable. To avoid the
systematic error induced by the small deviations of our $g_c$ value from the true critical point,
we here only use system sizes up to $L=64$. We have estimated that the deviations will then affect the extracted exponents
less than the purely statistical errors of the
fitting parameters. All estimated $\nu$ values for the four trial states are consistent with each other and with the known value of
the exponent. The results are listed in Tab.~\ref{table2}. We can see good agreement with the correct O(3) exponent in all cases.

For $a=0.5$, the data, shown in Fig.~\ref{sL}(b), show more dependence on the trial state, but in all cases the slope takes the expected
value for sufficiently large system sizes. For $a=0.125$, we can see very significant dependence on the trial state, but here as well the
slope eventually crosses over to the correct critical form for large $L$. The results for $a=0.5$ are also listed in Tab.~\ref{table2}, but
for $a=0.125$ we did not carry out the analysis in detail because of the small number of points falling in the asymptotic scaling regime.
Nevertheless, these tests make clear that there is a cross-over size, which increases with decreasing $a$ and depends on the trial state,
above which the critical O(3) scaling is obtained. The non-universal prefactor of the scaling function depends strongly on $a$ and
the trial state.

\begin{table}[b]
\caption{
  Exponent $\nu$ obtained from the $g$ derivatives of $\rho_s L$ and $Q$ at the best estimated critical point $g_c$. To minimize the systematic
  errors originating from the deviation of the $g$ value used from the true critical point, we only used system sizes up to $L=64$ in the fits
  giving the results shown here. The minimum size is indicated in each case.}
\begin{tabular}{|c|c|c|c|c|c|c|}
\hline
\ &\multicolumn{6}{|c|}{a=1}\\
\hline
\ trial state & $\nu(\rho_s)$ & $\chi^2(\rho_s)$ &  $L_{min}$ & $\nu(Q)$&$\chi^2(Q)$& $L_{min} $ \\
\hline
$\Psi_1$ & 0.705(7) & 1.3  & 24  & 0.716(7) & 1.5 & 16\\
\hline
$\Psi_2$ & 0.706(5)  & 1.4  & 24 & 0.718(7) & 1.3 & 24\\
\hline
$\Psi_3$ & 0.707(7) & 1.5  & 24  & 0.710(6) & 1.1 & 16 \\
\hline
$\Psi_4$ & 0.713(7)   & 1.3  & 28 & 0.711(5) & 0.9 & 16\\
\hline
\ &\multicolumn{6}{|c|}{a=0.5}\\
\hline
\  & $\nu(\rho_s)$ & $\chi^2(\rho_s)$ &  $L_{min}$ & $\nu(Q)$&$\chi^2(Q)$& $L_{min} $ \\
\hline
$\Psi_1$ &  0.704(20)      & 1.5 & 32 & 0.709(20)    & 0.9  &  28\\
\hline
$\Psi_2$ &  0.709(8)        & 0.9 & 20 & 0.709(8) & 1.4 &  12\\
\hline
$\Psi_3$ &  0.707(9)        & 0.8 & 28 & 0.714(9) & 0.8 &  24\\
\hline
$\Psi_4$ &  0.709(10)      & 1.3 & 24 & 0.708(9) & 1.1 &  16 \\
\hline
\end{tabular}
\label{table2}
\end{table}

\begin{figure}[t]
  \includegraphics[width=7cm]{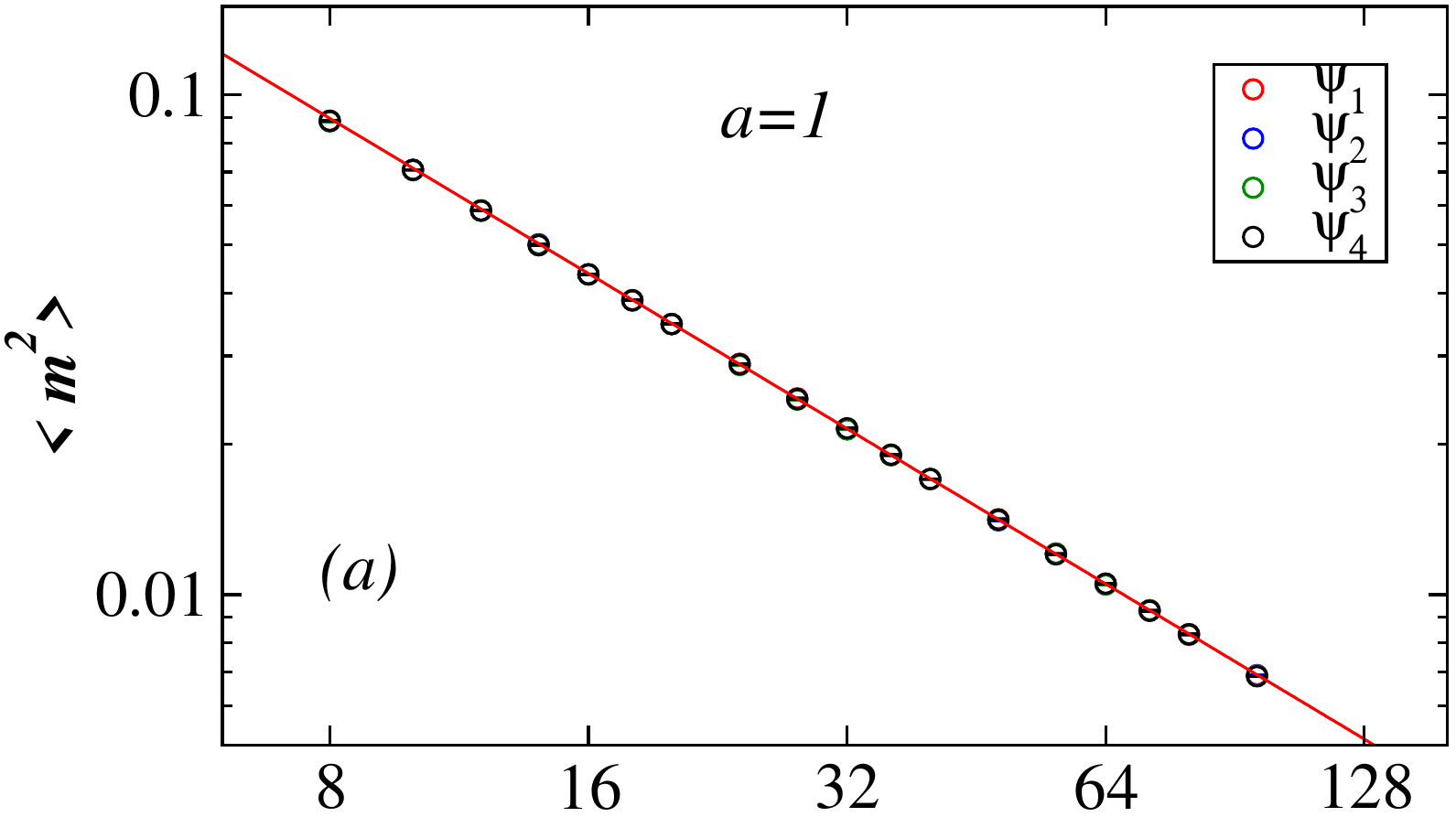}
  \vskip2mm
  \includegraphics[width=7cm]{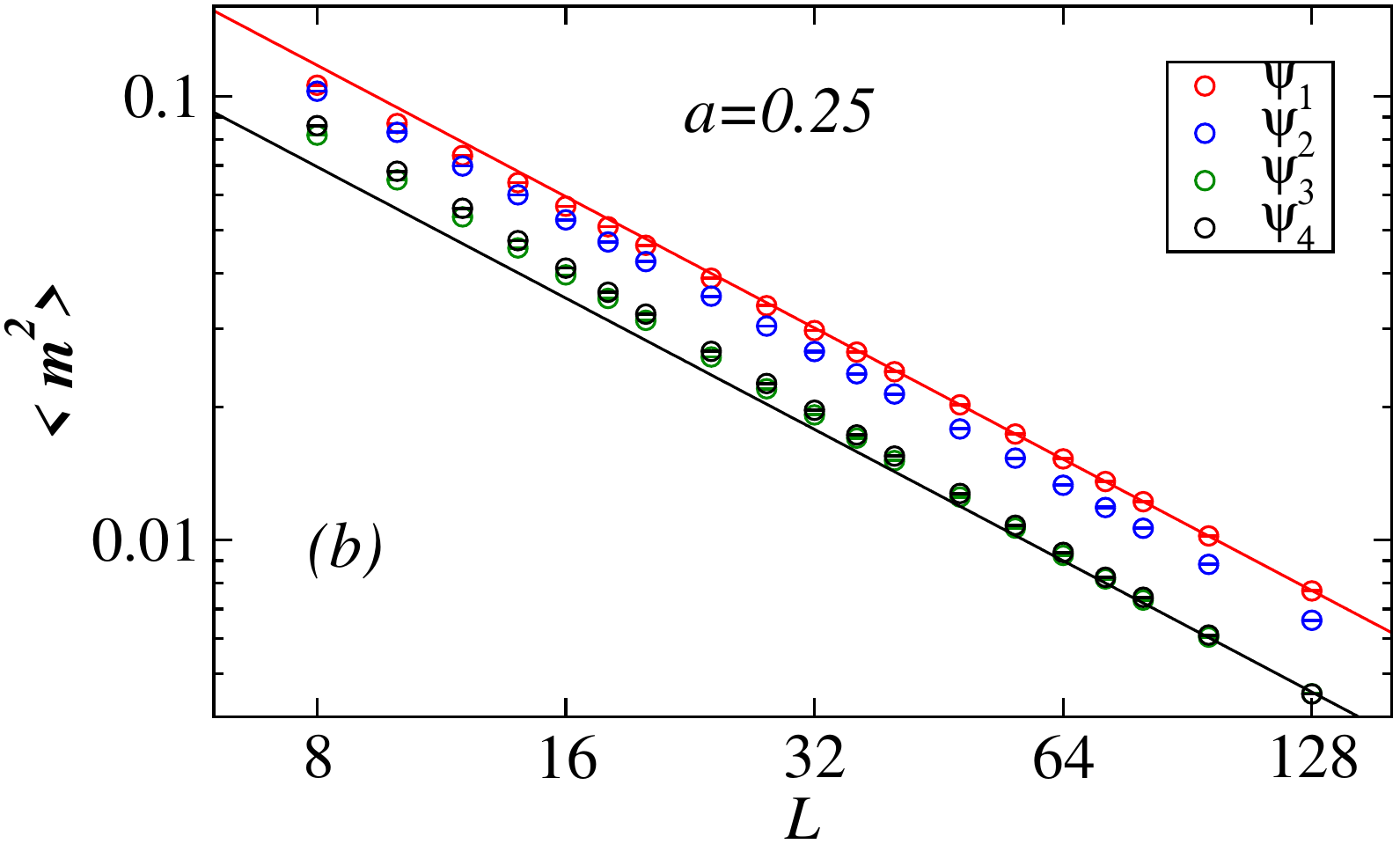}
 \caption{The squared order parameter at the best estimated critical point $g_c=2.5222$ versus the system size.
 for (a) $a=1$ and (b) $a=0.5$. The solid lines are fits to the large-$L$ data.}
\label{peta}
 \end{figure}

Next, we investigate the exponent $\eta$ (the anomalous dimension) of the critical correlation function. We use the squared staggered
magnetization $\la {\bf m}^2\ra$ at $g_c=2.5222$. In this case we include all our large system sizes in the fits, based on an estimation
of the effects of the precision of the $g_c$ values. For $a=1$, as shown in Fig.~\ref{peta}(a), the square order parameter $\la {\bf m}^2\ra $
scales well according to the expected critical form Eq.~(\ref{m2L}) with increasing size $L$, with only a very weak dependence on the
trial state. For the case $a=0.25$, significant differences in the values of $\la {\bf m}^2\ra$ can be observed for the
different trial states, as illustrated in
Fig.~\ref{peta}(b). Nevertheless, each group of data points from the same trial state forms a straight line on the double-log graph. For
large enough $L$ the correction to scaling vanishes and within statistical errors the slopes of the four lines are identical and fully
consistent with the known value of $\eta$. Fitting to the expected finite-size form, again leaving out system sizes smaller than $L_{\rm min}$
chosen such that the fits are acceptable, we obtain values listed in Tab.~\ref{teta} for $a=1$ and $0.5$. For $a=0.25$, the results are
also consistent with the O(3) exponent, but the statistical errors are much larger, due to the large $L_{\rm min}$ needed in this case,
and we do not list the results.

\begin{table}[b]
\caption{Exponent $\eta$ obtained from finite-size scaling of the staggered magnetization $\la {\bf m}^2\ra$ at the estimated critical
point $g_c$, using the different trial states and projection times. The O(3) value of the exponent is $\eta=0.0375(5)$. \cite{Vicari}}
\begin{tabular}{|c|c|c|c|c|}
\hline
\multicolumn{5}{|c|}{a=1}\\
\hline
 trial state & $\Psi_1$ & $\Psi_2$ & $\Psi_3$ & $\Psi_4$ \\
\hline
$\eta$      & 0.036(3) & 0.032(3) & 0.038(6) & 0.039(3)  \\
\hline
$\chi^2$   & 1.6       & 1.6      &1.8          & 1.3 \\
\hline
$L_{min}$& 48          &  24          & 18          &  40 \\
\hline
\multicolumn{5}{|c|}{a=0.5}\\
\hline
 trial state  & $\Psi_1$ & $\Psi_2$ & $\Psi_3$ &$\Psi_4$ \\
\hline
$\eta$       & 0.026(6) & 0.036(1) & 0.037(2) & 0.036(4)  \\
\hline
$\chi^2$    & 1.9       & 1.6       &1.9           &       0.81 \\
\hline
$L_{min}$ & 16          &  20         & 20           &         56\\
\hline
\end{tabular}
\label{teta}
\end{table}

\subsection{Critical Binder Ratio}

We have shown that typical critical fluctuations emerge out of arbitrary trial states when projecting in imaginary time
$\tau \propto L^z$, instead of projecting fully to the ground state of each finite system. However, as mentioned, from the
point of view of the path integral picture of the PQMC simulations, we can understand $a$ as time-space aspect ratio, with
the initial trial states corresponding to different kinds of boundary conditions. One can also think of this as a sudden
quench, where the Hamiltonian is changed at $\tau=0$ from the one (some times unknown one) for which the trial state is
the ground state to the critical Hamiltonian, followed by time evolution with the latter.

While the critical exponents are independent of the system geometry, the critical value of the Binder ratio $Q$ is universal
only in the sense that the value is determined by the dimensionality and symmetry of the system, irrespective of the details
of interaction and lattice structure, but under the condition that the boundary conditions and the geometry, e.g., aspect
ratio is fixed.\cite{KB,Todo} This implies that changing the temporal boundary conditions and the aspect ratio  should affect the
critical value of $Q$. Therefore it is expected that the critical value of $Q$ in the current PQMC simulations will
change with different trial states and factor $a$ in the time scaling $\tau = a L$. We investigate this dependence next

\begin{figure}[t]
  \includegraphics[width=7cm]{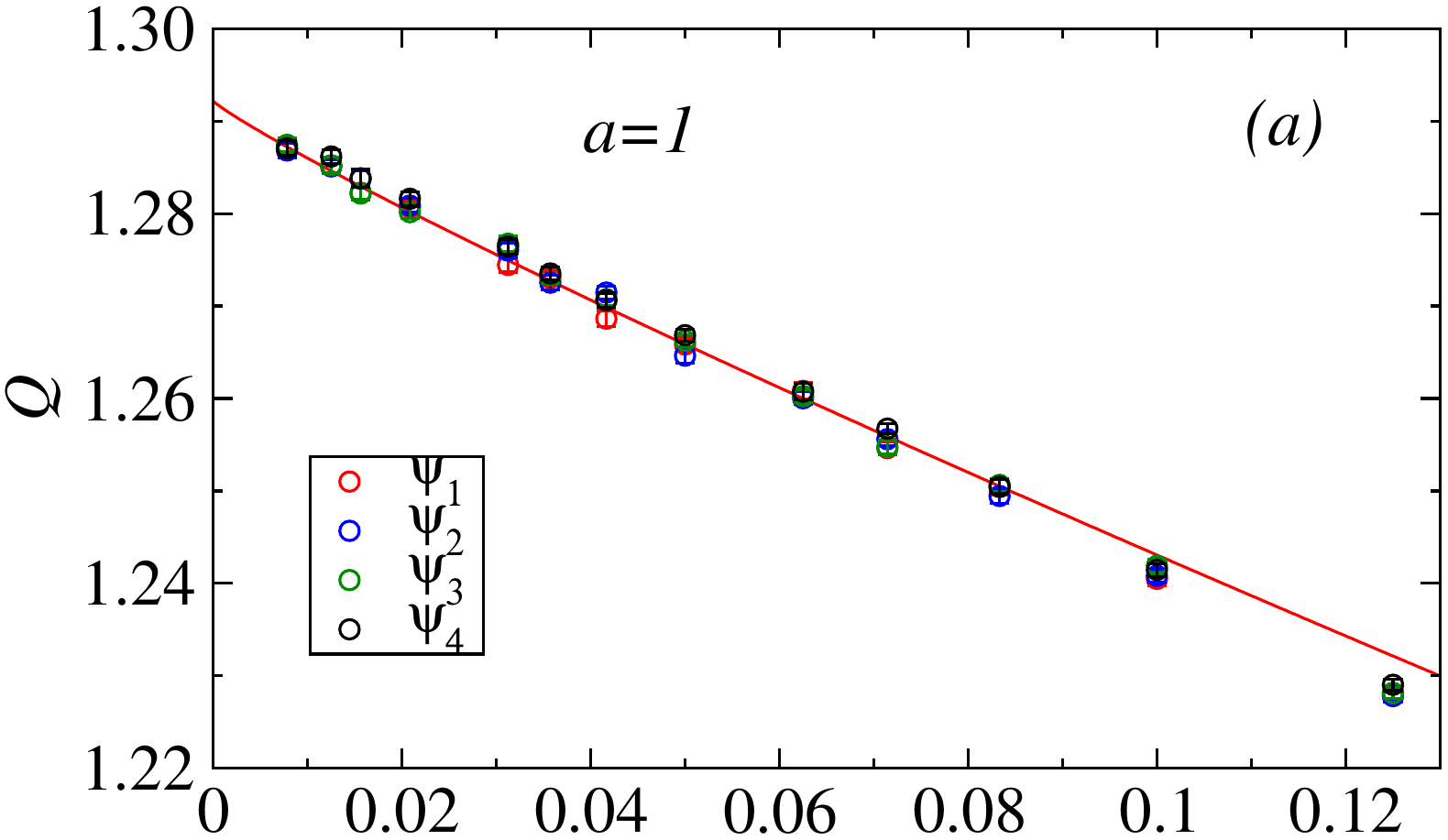}
  \vskip2mm
  \includegraphics[width=7cm]{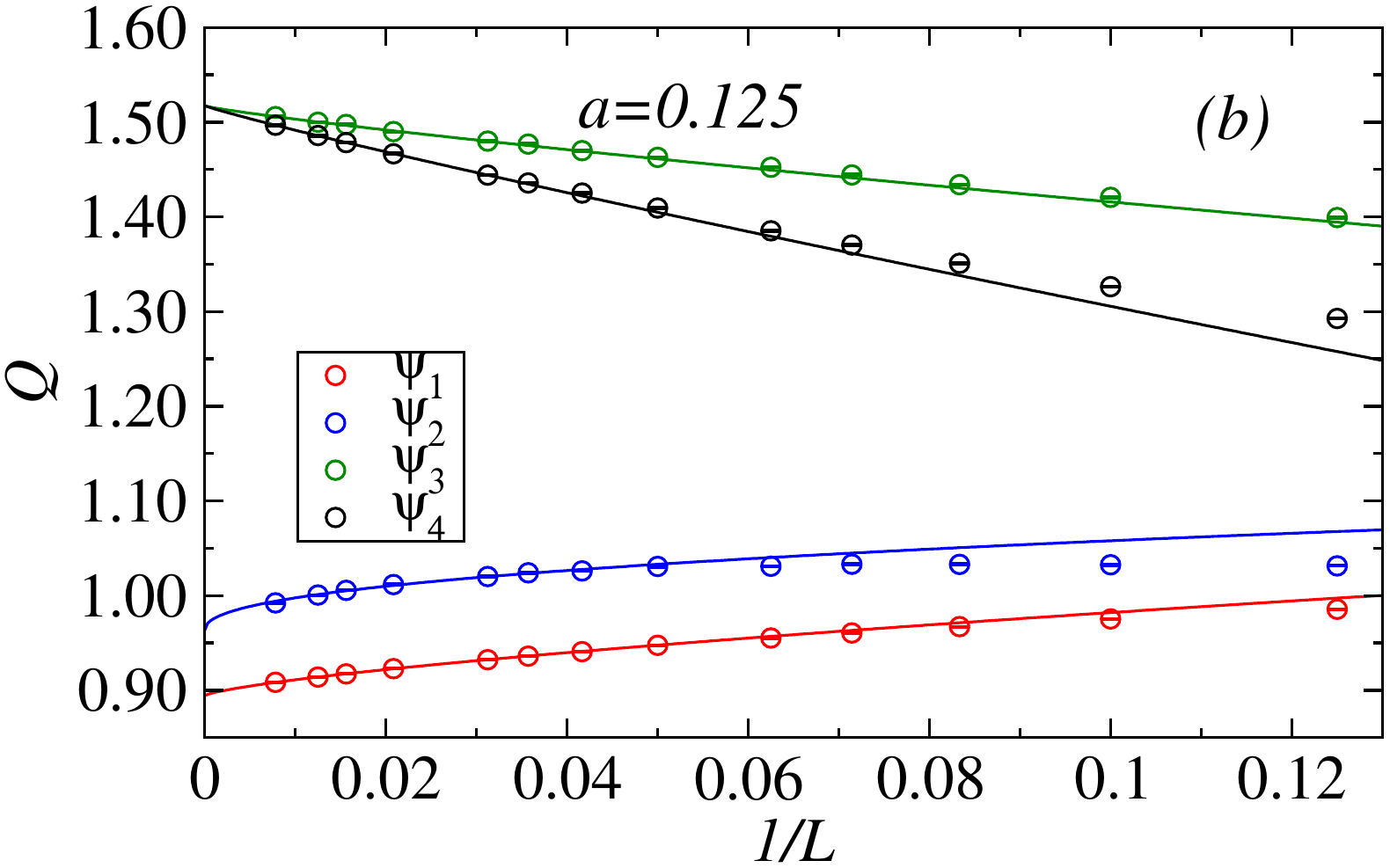}
\caption{Critical Binder ratios obtained in simulations with various trial states after projection time $\tau =a L$ with
  (a) $a=1$ and (b) $a=0.125$. The four curves are power-law fits used to extract the infinite-size values. In (a) we only show
  the fit to data from trial state $|\Psi_1\rangle$, for the sake of clarity.}
  \label{bindera}
 \end{figure}

Figure~\ref{bindera}(a) shows the Binder ratio at $g_c=2.5222$ versus $1/L$ in simulations with the four different trial
states and two aspect ratios. It is clearly seen that $Q_c$ changes dramatically when $a$ changes from $1$ to $0.125$
for each of the trial states. The value of $Q_c$ can be found by fitting Eq. (\ref{Ac}) to the data. However, for $a=1$, the
results for $Q_c$ obtained with all trial states
seem to converge to the same value when $L \to \infty$; for $|\Psi_1\ra$, $Q_c=1.290(3)$, with reduced $\chi^2=0.9$ and starting size
$L_{\rm min}=32$; for
$|\Psi_2\ra$, $Q_c=1.289(2)$ ($\chi^2=1.1$, $L_{\rm min}=20$), for $|\Psi_3\ra$, $Q_c=1.291(3)$ ($\chi^2=1.1, L_{\rm min}=20$); for
$|\Psi\ra_4$, $Q_c=1.289(2)$
($\chi^2=0.5, L_{\rm min}=32$). In these fits, the exponent $\omega$ in all cases is consistent with the known value
$\omega \approx 0.78$, though with relative statistical errors of about $20\%$ typically.
The extrapolated values of $Q$ agree well with $Q_c \approx 1.293(3)$ previously obtained for
an ``incomplete bilayer'' Heisenberg model (where the intra-layer couplings are missing on one layer), but differs slightly from
the result for the complete bilayer Heisenberg model, $Q_c=1.2858(3)$;\cite{dbAFH} most likely this disagreement
is due to an underestimated error bar in the previous calculation.

In the case of small $a=0.125$, it is clearly seen in Fig.~\ref{bindera}(b) that the differences in $Q_c$ obtained with
different trial state can be  drastic. Nevertheless, $Q_c=1.519(3)$ from  $|\Psi_3\ra$ and $Q_c=1.523(4)$ from $|\Psi_4\ra$
are still very similar; perhaps even identical asymptotically.
(The exponent $\omega$ found in the two fits agree well with 0.78.)
This may seem surprising, since these trial states are quite different and have different variational
energies. The trial states are similar in the sense that they are simple product states of singlets on neighboring sites,
but in one case the translational symmetry is broken and in one case it is not. One may speculate that the value of the
Binder ratio for small $a$ and $L \to \infty$ is related to the entanglement structure of the trial state. This would be
very interesting and deserves further study.

\begin{figure}[t]
  \includegraphics[width=7.5cm]{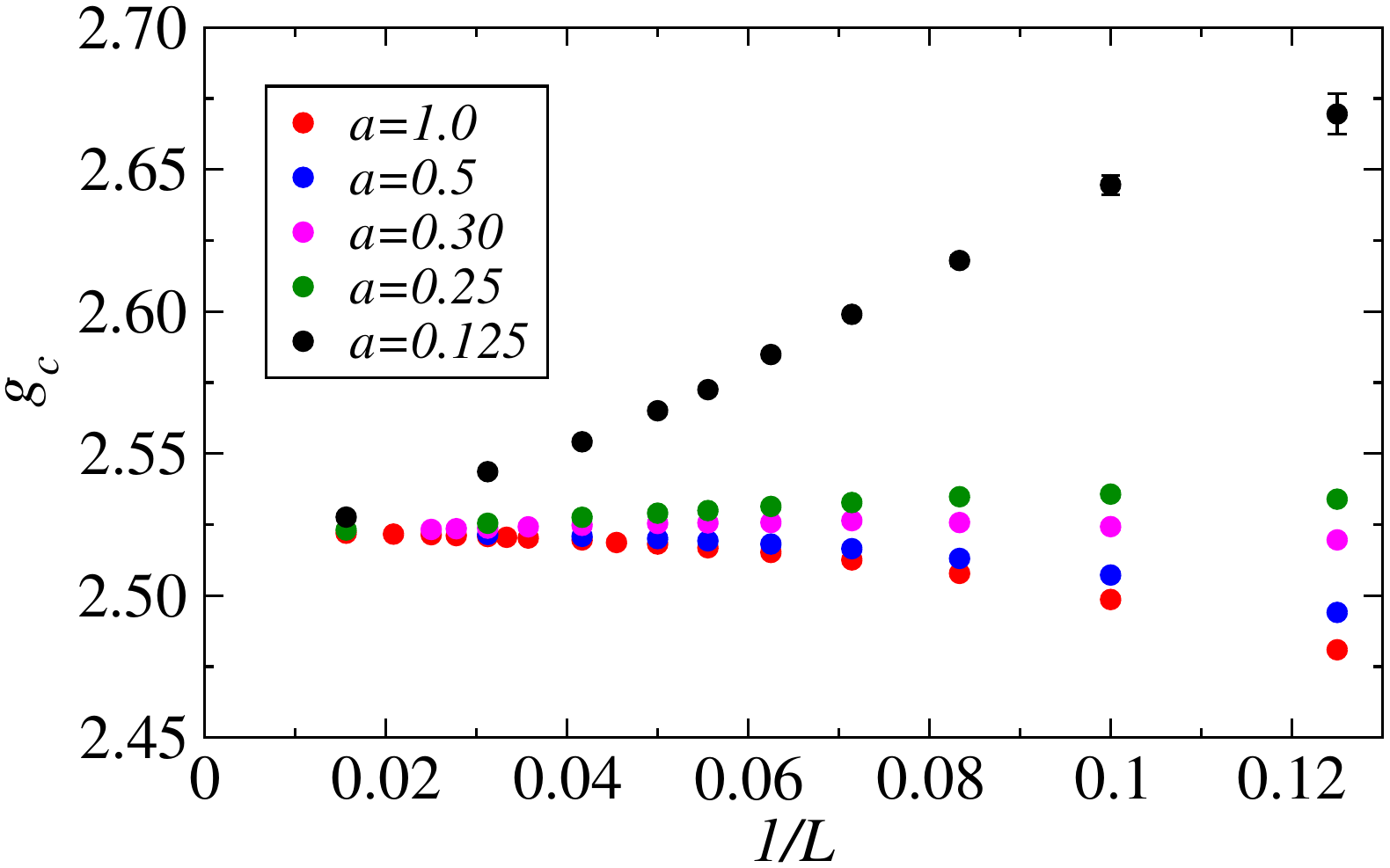}
  \caption{Crossing point between Binder cumulants for system sizes $(L,2L)$ graphed versus $1/L$ for different time
    scalings $\tau=aL$ with the trial state $|\Psi_1\ra$.}
  \label{psi1fig}
 \end{figure}

\section{Discussion and conclusions}
\label{conclusion}

We have considered the concept of typicality to quantum critical points approached in PQMC simulations, where an initial
state is, in effect, subject to an instantaneous quench followed by imaginary-time evolution with the critical Hamiltonian.
The initial (trial) states can be thought as different temporal boundary conditions. Typicality here corresponds to an
insensitivity of the universal critical fluctuations of the projected (evolved) state to the details of the initial
state---even for trial states that are very poor in the variational sense---when the time $\tau$ of the evolution scales
as $\tau = aL$.

By studying the bilayer Heisenberg model as an example, we have confirmed that the correct quantum-critical exponents are
reproduced for a range of different trial states (supporting a complete independence on the trial state) and arbitrary factors $a$,
after some cross-over system sizes that increases for decreasing $a$. While the correct critical exponents are always obtained
for sufficiently large $L$, various non-universal numbers depend strongly on $a$ even for $L \to \infty$.

While the typicality in the above sense is not too surprising, considering the similarity with $T>0$ simulations where it is well known
that one can scale the inverse temperature $\beta=bL^z$ to study quantum-critical scaling with the independent variable $\beta$ eliminated,
the freedom of choosing the trial state in projector simulations goes beyond the $T>0$ formalism. Our purpose here has been
to confirm the typicality for a wide range of trial states, and also to make some observations that may be useful in practice.
Beyond PQMC simulations, the typicality may also be very useful in calculations with tensor
network states, where projection out of an initial state is often done in order to optimize the ground state. For studies
of quantum critical points, it should be sufficient to project out to $\tau \propto L^z$ if $z$ is known, and if $z$ is not
known it should also be possible to extract its value by studying the dependence of results on $\tau$.

Naively, one might expect that it should always be better to choose a large factor $a$, so that the true ground state is projected out.
However, our results, e.g., in Fig.~\ref{crossing}, reveal that the leading finite-size scaling corrections, i.e., those governed by the
exponent $\omega$, can change sign as $a$ is varied. This indicates the interesting and practically useful possibility that the
leading corrections actually vanish at some special value of $a$. To make this point clearer, in Fig.~\ref{psi1fig}, we show results obtained
with the trial state $|\Psi_1\ra$ and several values of $a$. Here we can see the change in sign of the correction very clearly,
and it appears that $a\approx 0.3$ is the optimal value for canceling the leading correction. The corrections at this point would then
be governed by the following correction exponent $\omega_2$. Since the amplitudes of the various scaling corrections are
non-universal and also vary between different quantities, one may have to optimize $a$ for each quantity of interest in order
to take advantage of this effect, and for some quantities one may not even be able to find such an optimal values (and this
may also depend on the trial state used). Thus, the method of optimizing the projection time may not be quite as powerful
as the well known method of tuning an interaction in a model to reach a point at which the leading correction is absent for
all quantities.\cite{Vicari} Nevertheless, this effect may potentially be very helpful in some cases. Moreover,
apart from finding points where corrections are vanishing,
it may also be useful in finite-size scaling studies to do common fits to results for several different values of $a$, since
the mix of leading and subleading corrections depend strongly on $a$ and the fits may become more stable with such information
present in the data set. We are planning to explore these issues further in both PQMC and $T>0$ simulations (where optimal
prefactors $b$ of the inverse temperature, $\beta=bL^z$, may exist).

\acknowledgments
This work is supported by the National Natural Science Foundation of China under Grant No.~11734002
and 11775021 (W.G.), by the National Science Foundation under Grant No.~DMR-1710170 (A.W.S.) and by a Simons Investigator Award (A.W.S.).


\begin{thebibliography}{99}
\bibitem{mlnote}
We note that it was claimed in Ref.~\onlinecite{carleo17}
that a wave function optimized and sampled by machine learning teachniques can produce better
energies than previously obtained variational energies for the 2D Heisenberg model. However,
the far better variational results obtained in Ref.~\onlinecite{lin12} were not included in the comparison.
\end{thebibliography}
\end{document}